\documentclass[prl,twocolumn,10pt,aps,superscriptaddress]{revtex4-1}
\usepackage{amsmath,amssymb,bm}
\usepackage{graphicx}
\usepackage{epstopdf}
\usepackage{latexsym}
\usepackage{subfigure}
\usepackage{color}
\usepackage{natbib}
\usepackage{hyperref}
\usepackage{braket}
\usepackage{dsfont}
\usepackage[english]{babel}
\usepackage{blindtext}
\hypersetup{
  colorlinks,
  citecolor=magenta,
  linkcolor=blue,
  urlcolor=blue}
      
\begin{document}

\title{Entanglement entropy and deconfined criticality: \\ emergent SO(5) symmetry and proper lattice bipartition}

\author{Jonathan D'Emidio}
\email{jonathan.demidio@dipc.org}
\affiliation{Donostia International Physics Center, P. Manuel de Lardizabal 4, 20018 Donostia-San Sebasti\'an, Spain}
\author{Anders W. Sandvik}
\email{sandvik@bu.edu}
\affiliation{Department of Physics, Boston University, 590 Commonwealth Avenue, Boston, Massachusetts 02215, USA}
\affiliation{Beijing National Laboratory for Condensed Matter Physics, Institute of Physics, Chinese Academy of Sciences, Beijing 100190, China}

\begin{abstract}
We study the R\'enyi entanglement entropy (EE) of the two-dimensional $J$-$Q$ model, the emblematic quantum
spin model of deconfined criticality at the phase transition between antiferromagnetic and 
valence-bond-solid ground states. State-of-the-art quantum Monte Carlo calculations of the EE reveal 
 critical corner contributions that scale logarithmically with the system size, with a coefficient in remarkable 
agreement with the form expected from a large-$N$ conformal field theory with SO($N=5$) symmetry. However, 
details of the bipartition of the lattice are crucial in order to observe this behavior. If the subsystem 
for the reduced density matrix does not properly accommodate valence-bond fluctuations, logarithmic contributions 
appear even for corner-less bipartitions. We here use a $45^\circ$ tilted cut on the square lattice. Beyond 
supporting an SO($5$) deconfined quantum critical point, our results for both the regular and tilted cuts 
demonstrate important microscopic aspects of the EE that are not captured by conformal field theory.
\end{abstract}
\maketitle

The bipartite entanglement entropy (EE), a measure of the amount of entanglement for a given wavefunction bipartition, has become an important characteristic of quantum phases of matter and their phase transitions
\cite{Vidal03,Calabrese04,Amico06,Laflorencie16}. The dominant contribution to the EE of a $d$-dimensional system grows as $l^{d-1}$, where $l$ 
is the length of the subsystem used to define the reduced density matrix. The corrections to this ``area law'' \cite{Srednicki93} contain important 
information on the nature of the ground state of the system, e.g., the constant $\gamma$ characterizing topological order 
\cite{Kitaev06,Levin06} and the logarithmic correction to the area law that originates from gapless Goldstone modes in systems breaking 
O($N>1$) symmetry \cite{Song11,Kallin11,Metlitski2011}. Logarithmic corrections also appear in critical spin chains \cite{Vidal03,Korepin04}, but not at 
criticality for $d=2$ when the bipartition boundary is smooth (corner-less). According to conformal field theory (CFT), logarithmic corrections 
do arise from corners in two-dimensional (2D) quantum-critical systems \cite{Fradkin06,Casini07}. These corner contributions are potentially useful for 
establishing the proper CFT description of a given quantum phase transition studied numerically 
\cite{Singh12,Kallin13,Kallin14,Helmes14,Helmes15}, or, conversely, to rule out a CFT description.

We here investigate corner logarithms at a putative deconfined quantum-critical point (DQCP) \cite{Senthil04a,Senthil04b,Levin04}, 
computing the second R\'enyi EE using quantum Monte Carlo (QMC) simulations of the most well studied $S=1/2$ spin 
model exhibiting a transition between a N\'eel antiferromagnetic (AFM) and a valence-bond-solid (VBS) ground state; the $J$-$Q$ 
model \cite{Sandvik07}. Our results provide further insights into the SO($5$) nature of the AFM--VBS phase transition and highlight
the role of microscopic (lattice) details of the bipartition.

{\em Conundrum of EE Anomalies.}---The CFT prediction for the EE of a critical system with corners is \cite{Fradkin06,Casini07}
\begin{equation}
S(l) = \mathcal{A}l - a\ln{l} + \text{constant},
\label{sform}
\end{equation}
where $\mathcal{A}$ and the constant term are not universal, but  $a$ depends in a universal way (for a system with a given CFT description) on the number of corners 
(and their angles) of the subsystem. For any CFT, it is believed that $a>0$ \cite{Hirata07}.

An apparent violation of the positivity requirement of $a$ was recently invoked \cite{Zhao22,Liao23,Song23a} to argue that AFM--VBS transitions in 2D 
SU($N$) quantum magnets with small $N$ are not described by CFTs. $J$-$Q$ type models exhibiting AFM--VBS transitions for any $N$ \cite{Lou09,Kaul12} 
were studied using QMC simulations, and only above $N=7$ does $a$ turn positive \cite{Song23a}. Later work \cite{Song23b} seemed to show that the 
EE even for smooth subsystems has logarithmic corrections. Potentially, these anomalies could reflect the Goldstone modes expected if the AFM--VBS 
transition is weakly first-order \cite{Jiang08,Chen13,Demidio23a}, as it is now expected to be in all quantum spin models studied so far
\cite{Takahashi23} and likely also in the related classical 3D loop model \cite{Nahum15a,Nahum15b}. 

Indeed, in a recent study \cite{Deng23a} of the $J$-$Q_3$ model with a smooth bipartition, an EE scaling form with finite-size corrections \cite{Deng23b} 
delivered a value of $a$ in Eq.~(\ref{sform}) corresponding to four Goldstone modes, likely reflecting close proximity to an expected \cite{Nahum15b} 
critical point with emergent SO(5) symmetry. Universal near-criticality is also supported by past works on scaling behavior, which have found 
consistency between critical exponents extracted in different ways and in different models \cite{Sandvik12,Nahum15a,Sandvik20}. The question we 
ask in this work, and answer in the affirmative, is whether the logarithmic corner contributions expected for an SO(5) CFT can also be observed, 
contrary to the claims\cite{Zhao22,Liao23,Song23a,Song23b} of no underlying CFT. 

\begin{figure}[t]
\centerline{\includegraphics[angle=0,width=0.95\columnwidth]{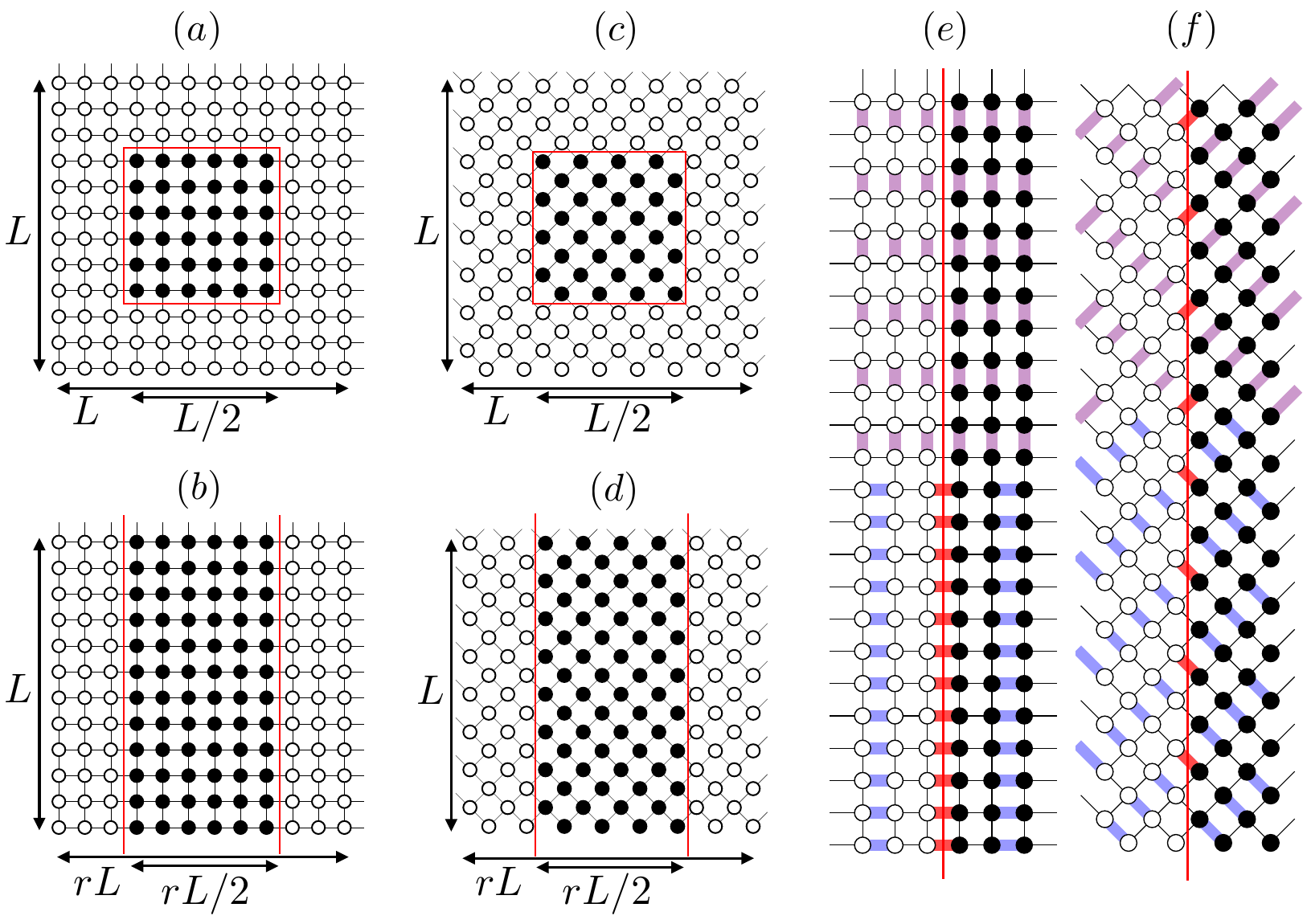}}
\vskip-2mm
\caption{Lattices and EE subsystem cuts considered in this work. The full systems all have periodic boundary conditions in both horizontal and 
vertical directions. (a), (b): The standard orientation of the lattice links, with the lattice constant taken as unity.  In (a), the system
size is $L^2$ and the square cut has length $L/2$. For the smooth cut in (b), the vertical length is $L$ and the horizontal length is $rL$
with $r=1,2$. (c), (d): $45^\circ$ tilted lattices, where we take the lattice constant as $1/\sqrt{2}$ and the number of spins is $2rL^2$. 
The EE subsystem in (c) contains $L^2/2$ spins (for even $L$) and in (d) half of all spins. (e), (f): Illustration of VBS patterns at the subsystem
edge. With the straight cut in (e), one of the VBS patterns is disfavored over the other, since it contains a line of cut singlets shown in red.  
With the tilted cut in (f), all of the VBS patterns are treated equally, with the same density of cut singlets in each case.}
\label{fig:cuts}
\end{figure}

We argue that the standard way of cutting out the EE subsystem from the lattice is not ideal for a system with critical fluctuations of emergent 
degrees  of freedom that are not point-like. In a VBS, the objects that order are singlets (dimers) on nearest-neighbor lattice sites; 
on the square lattice forming four degenerate columnar patterns that break $Z_4$ symmetry. In the theory of DQCPs \cite{Senthil04a,Senthil04b,Levin04}, 
it is posited that the VBS fluctuations develop a U(1) symmetry as the AFM transition is approached, which indeed was found in several variants of the $J$-$Q$ 
model \cite{Sandvik07,Jiang08,Lou09}. Given that the reduced density matrix used to define the EE is related to an effective boundary Hamiltonian 
\cite{Li08}, it appears plausible to us that a requirement for the bipartition is that the boundaries of the subsystem  must equivalently treat all critical 
fluctuations that form out of the melted VBS state. Ideally, the subsystem should be defined in such a way as to allow 
tunneling between local VBS patterns in a way reflecting the emergent U(1) symmetry at the transition.

Using QMC simulations with a highly efficient method for extracting the second R\'enyi EE (described below), we will show that, it is 
only when the above requirement is fulfilled that the correct sign of the corner coefficient is obtained and the corrections from Goldstone modes are 
completely suppressed up to the largest system sizes studied here. We consider the standard square lattice of size $rL\times L$ as well as $45^\circ$ 
tilted variants (as used in Ref. \cite{Liao23}), with periodic boundary conditions and aspect ratios $r=1,2$. Within these systems we define smooth and cornered subsystems, as 
illustrated in Fig.~\ref{fig:cuts}. It is easy to see that the standard square cuts in Figs.~\ref{fig:cuts}(a) and \ref{fig:cuts}(b) violate our proposed 
requirement, while it is satisfied with the cuts of the tilted lattices in Figs.~\ref{fig:cuts}(c) and \ref{fig:cuts}(d). As shown in Figs.~\ref{fig:cuts}(e) 
and \ref{fig:cuts}(f), certain VBS configurations can be favored or disfavored by the subsystem boundary 
in the case of the standard square lattice cut.  However, for the tilted system, the boundaries treat the four different VBS configurations 
equally, with the same density of cut singlets in each case. Accordingly, only the tilted cut ensures that the reduced density matrix can capture 
correctly the quantum fluctuations of the subsystem. Indeed, in this case we find the expected sign of the corner coefficient and no discernible
logarithmic corrections to the area law in the absence of corners.

Remarkably, with the tilted subsystem, we find that the corner coefficient matches essentially perfectly the form predicted for an O($N$) CFT with large 
$N$, for the specific value $N=5$.  Although exact agreement with the O($N$) theory is not expected, this large corner coefficient is strong evidence for an enlarged number of degrees of freedom at the DQCP.  An emergent SO(5) symmetry, which in the present context is the same as O($5$), indeed is imposed in one class of 
DQCP theories \cite{Tanaka05,Senthil06,Nahum15b}. This symmetry was first detected in a 3D loop model \cite{Nahum15a,Nahum15b} and later also in a variant 
of the $J$-$Q$ model \cite{Takahashi20}.  Moreover, a multicritical SO($5$) point studied using the numerical conformal bootstrap method \cite{Chester23} lends 
support to a fine-tuned transition. Our results provide evidence of the AFM--VBS transition in the $J$-$Q$ models being extremely close to this multicritical 
CFT, despite ultimately exhibiting a weakly ordered coexistence state.

{\em Model and methods.}---Before presenting our numerical results to support the above claims, we define the $J$-$Q$ model and explain
an efficient scheme for evaluating the second R\'enyi EE within the stochastic series expansion (SSE) \cite{Sandvik10a} QMC method.

We focus on the simplest, SU(2) invariant, spin-$1/2$ square-lattice $J$-$Q$ model \cite{Sandvik07}, with the Hamiltonian 
\begin{equation}
\label{eq:Hjq}
H= J\sum_{\langle ij \rangle}(\vec{S}_{i} \cdot \vec{S}_{j} -\tfrac{1}{4}) 
- Q\hskip-1mm\sum_{\langle ijkl \rangle}(\vec{S}_{i} \cdot \vec{S}_{j} -\tfrac{1}{4})(\vec{S}_{k} \cdot \vec{S}_{l} -\tfrac{1}{4}),
\end{equation}
where $J>0$ is the AFM nearest-neighbor coupling and the four-spin interactions ($Q>0$) is a product of two adjacent singlet 
projectors, including both $\hat{x}$ and $\hat{y}$ orientations. These interactions, with the negative sign, induce locally correlated 
singlets, and the ground state  undergoes a transition from an AFM to a VBS as $Q/J$ is increased. Many works have studied quantum-critical 
scaling in the $J$-$Q$ model in the context of the still controversial DQCP proposal 
\cite{Sandvik07,Lou09,Jiang08,Chen13,Demidio23a,Melko08,Sandvik10b,Block13,Harada13,Suwa16,Shao16,Ma18,Takahashi20}. The most likely 
scenario based on recent works \cite{Sandvik20,Zhao20a,Demidio23a,Takahashi23} is that the transition is first-order but with a very weakly
ordered AFM--VBS coexistence state that, up to large length scales, retains the emergent SO(5) symmetry \cite{Nahum15b,Takahashi20} 
and critical fluctuations governed by meaningful exponents. Here we use the best current estimate $(J/Q)_c=0.04502$ \cite{Takahashi23} of 
the transition point and perform SSE QMC simulations at  temperature $T = 1/L$ (in units with Q=1).  We checked that our logarithmic terms are free from finite temperature effects \cite{Supp}.

While the von Neumann EE is out of reach of QMC methods, the R\'enyi EEs can be computed.   Here we focus on the second R\'enyi EE for a subsystem $A$ of the entire lattice $A \cup \bar A$, defined as 
$S_2=-\ln(\text{Tr}(\rho_A^2))$, where the reduced density matrix involves tracing over the spins in $\bar A$ and
$\rho_A = \text{Tr}_{\bar A}(e^{-\beta H})/Z$. 

Early QMC approaches for the R\'enyi EE of spin systems were based on the swap operator in valence-bond ground-state QMC \cite{Hastings10}. Later, the particular trace operation corresponding to the $n$\textsuperscript{th} R\'enyi entropy (often called the ``replica trick") \cite{Calabrese04,Buividovich08} was also adapted to SSE \cite{Melko10,Isakov11,Humeniuk12,Kulchytskyy15}.  These methods allowed for studies of many relevant systems, although it remained difficult to achieve small error bars for large systems.  It was only with the recent application of nonequilibrium work relations in the context of the R\'enyi EE put forward in Ref.~\cite{Demidio20}, originally inspired by Alba  \cite{Alba17}, that precise calculations with very large lattice sizes (more than $10^4$ spins) became possible.  The approach that we implement here within the SSE is an equilibrium variant of this method, which is simpler and more flexible, and in fact was recently introduced in the context of fermionic systems \cite{Demidio23b}.

We first express $S_2$ as a ratio of partition functions (free-energy difference): $e^{-S_2} =Z_A / Z_{\varnothing}$, where
$Z_A \equiv  \text{Tr}_{A}[(\text{Tr}_{\bar{A}}(e^{-\beta H}))^2]$, and $\varnothing$ denotes the empty set (no subsystem).  
Notice that $Z_{\varnothing}= Z^2$, where $Z=\text{Tr}(e^{-\beta H})$ is the partition function.  The powerful method 
that we leverage here for computing the ratio $Z_A / Z_{\varnothing}$ relies on the fact that one can find an interpolating ensemble 
$\mathcal{Z}(\lambda)$ where (1) the endpoints are given by $\mathcal{Z}(0)=Z_{\varnothing}$ and $\mathcal{Z}(1)=Z_{A}$ and (2) 
ratios of the form  $\mathcal{Z}(\lambda_i)/\mathcal{Z}(\lambda_j)$ can be computed extremely efficiently.

A very useful distribution satisfying these criteria was given in Ref.~\cite{Demidio20};
\begin{equation}
\label{eq:Zlam}
\mathcal{Z}(\lambda)=\sum_{C \subseteq A} \lambda^{N_C} (1-\lambda)^{N_A - N_C} Z_C,
\end{equation}
where the sum over $C \subseteq A$ includes all proper subsets of the set of spins in $A$, i.e., a number $0 \le N_C \le N_A$ of any of the
spins drawn from the $N_A$ spins within $A$, including the empty set $\varnothing$ and $A$ itself. Clearly property (1) is satisfied, and (2) can be 
computed via reweighting as
\begin{equation}
\label{eq:Zrat}
\frac{\mathcal{Z}(\lambda_j)}{\mathcal{Z}(\lambda_i)}=\left\langle \left(\frac{\lambda_j}{\lambda_i}\right)^{N_C} \left(\frac{1-\lambda_j}{1-\lambda_i}\right)^{N_A-N_C} \right\rangle_{\lambda_i}
\end{equation}
which is computed in an {\em equilibrium} QMC simulation of the $\mathcal{Z}(\lambda)$ ensemble with $\lambda=\lambda_i$. During the simulation spins within the fixed $A$ subregion are added to and removed from the fluctuating set $C$ according 
to detailed balance, in addition to the QMC updates; here of the SSE method described in detail in Ref.~\cite{Sandvik10a}.

The power of this approach lies in the interpolating distribution in Eq. (\ref{eq:Zlam}) that gives a pathway from $\lambda: 0 \to 1$ along which the total entropy can be computed precisely for very large systems. This can be done in a number of ways, including direct estimation of the exponentially small ratio $Z_A/Z_{\varnothing}=\mathcal{Z}(1)/\mathcal{Z}(0)$ using long nonequilibrium quenches \cite{Demidio20,Deng23b,Deng23a}, or by breaking the ratio into a product of incremental ratios, of the form in Eq.~(\ref{eq:Zrat}), which can be computed either with equilibrium simulations using numerical integration \cite{Block20} or with short nonequilibrium quenches \cite{Zhao22,Song23a,Song23b}.  The latter approaches are akin to  “ratio tricks” employed early on with the swap operator \cite{Hastings10}  in the ground state and with the replica trick at finite temperature \cite{Humeniuk12,Kulchytskyy15}.

The approach that we employ here also makes use of the ratio trick by introducing many intermediate equilibrium simulations at different values of $\lambda$.  However, this approach is significantly simpler than the nonequilibrium methods (which require frequent re-equilibration of starting configurations and fixing a quench time in advance) and is also free from numerical integration errors, owing to the statistically exact Eq.~(\ref{eq:Zrat}).  In general we find that schemes making use of Eq. (\ref{eq:Zlam}) are at least an order of magnitude more efficient than traditional methods. They all have the same efficiency within appropriate limits \cite{Supp}.

\begin{figure}[t]
\centerline{\includegraphics[angle=0,width=0.925\columnwidth]{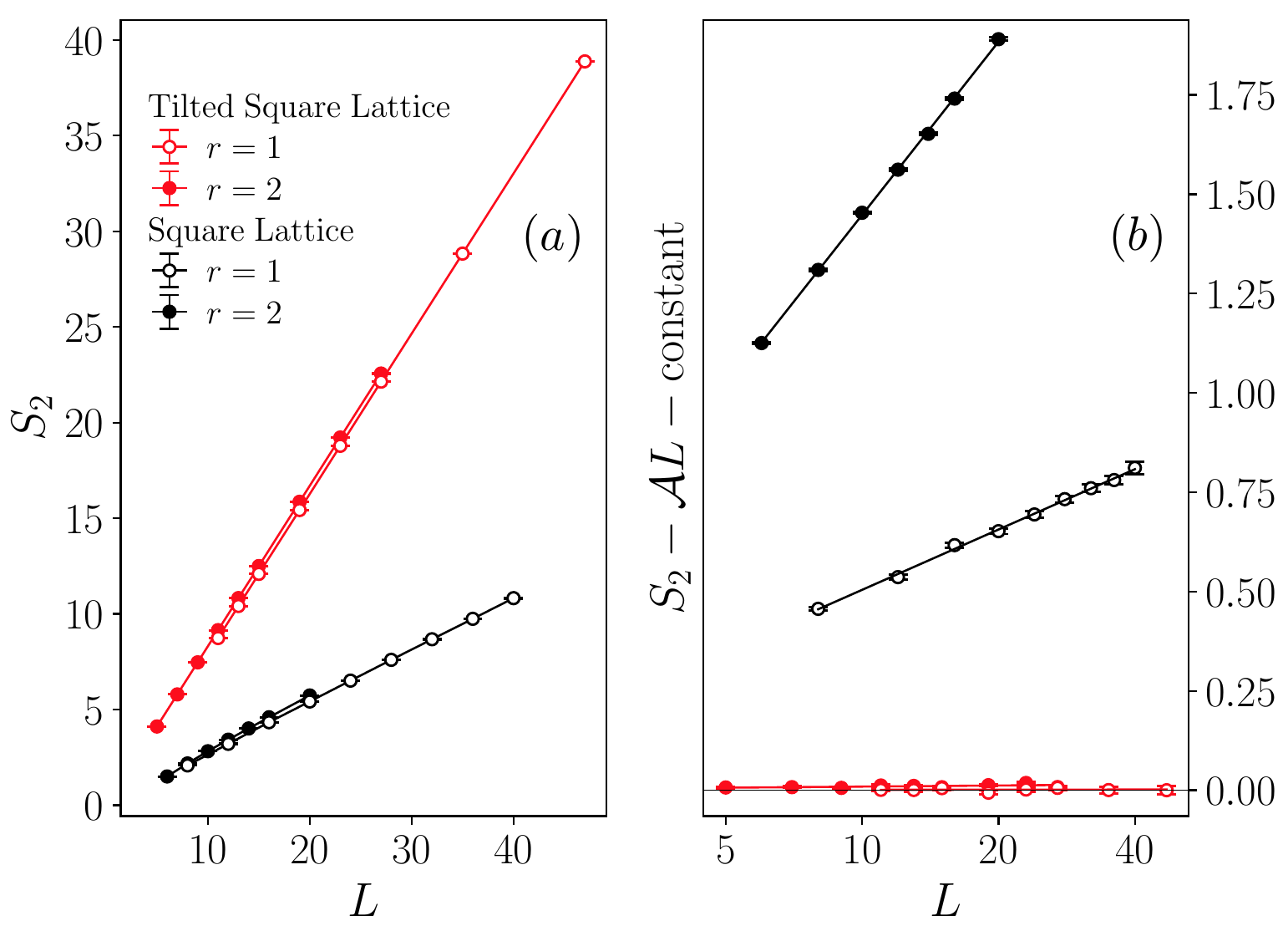}}
\vskip-2mm
\caption{(a) Size dependence of the EE on $rL\times L$ lattices for $r=1,2$ and smooth cuts of the torus, with the subsystem containing half of the 
spins in each case. The results have been fitted to Eq.~(\ref{sform}), including the logarithmic correction that is not expected in a critical
system. (b) The same results with the fitted area law and constant contributions subtracted off, using a logarithmic $L$ scale.  The square lattices 
give $a=-0.22(2)$ and $a=-0.63(3)$ for $r=1,2$, respectively, while the tilted lattices give $a=0.00(2)$ and $a=0.004(8)$.  The full fits shown here are $S_2=0.262(1)L+0.22(2)\log(L)-0.47(3)$ and $S_2=0.248(2)L+0.63(3)\log(L)-1.11(4)$ for $r=1$ and $r=2$ of the square lattice, and $S_2=0.8372(8)L+0.00(2)\log(L)-0.47(4)$ and $S_2=0.8384(6)L+0.004(8)\log(L)-0.08(1)$ for $r=1$ and $r=2$ of the tilted square lattice.}
\label{fig:f2}
\end{figure}

%jqEE_tiltedcylinderoddBL
%0.8371699586553751
%0.0004638985960359596.     0.8372(8)L+0.00(2)\log(L)-0.47(4)
%-0.4699813768842247
%[0.00084993 0.01814621 0.03638025]
%[11. 13. 15. 19. 23. 27. 35. 47.]
%jqEE_rectiltedcylinderBL
%0.8383690431626375
%0.003975361305063403    0.8384(6)L+0.004(8)\log(L)-0.08(1)
%-0.08163803713070066
%[0.00066224 0.0080846  0.01138784]
%[ 5.  7.  9. 11. 13. 15. 19. 23. 27.]
%jqEE_sqlattcylinderBL
%0.2621104264006138
%0.2190505122833291.             0.262(1)L+0.22(2)\log(L)-0.47(3)
%-0.47115655705614223
%[0.00104701 0.01826533 0.03213472]
%[ 8. 12. 16. 20. 24. 28. 32. 36. 40.]
%jqEE_rectsqlattcylinderBL
%0.24764454253686594
%0.628899031114644.             0.248(2)L+0.63(3)\log(L)-1.11(4)
%-1.1059186744679321
%[0.00260812 0.02767206 0.03625692]
%[ 6.  8. 10. 12. 14. 16. 20.]

{\em Results.}---We begin by considering smooth cuts of $rL\times L$ systems. EE results for the both the standard and tilted lattice orientations, 
each with aspect ratios $r=1,2$, are shown in Fig.~\ref{fig:f2}(a) along with fits of the $L$ dependence to the three-parameter form in Eq.~(\ref{sform}).
In Fig.~\ref{fig:f2}(b) we show results with the dominant area law contribution and constant (i.e., those of the full fitted curve) subtracted off, using
a logarithmic size scale. 

In the case of the standard lattice orientation, there is a logarithmic contribution left, as also found previously \cite{Song23b,Deng23a}, and the 
coefficient increases by almost a factor of three when the aspect ratio is changed from $r=1$ to $r=2$; the results for $r=4$ are similar to $r=2$ and have 
therefore not included them in the graph. For the $r=1$ lattice the logarithm agrees with the previous results \cite{Zhao22,Song23a,Song23b,Deng23a}. 
In sharp contrast, for the tilted lattice there are no logarithmic contributions within statistical errors (see \cite{Supp} for a discussion of subtle even/odd effects).
We note that the area law term is expected to strongly depend on details of the lattice, including the angle of the subsystem boundary relative to the lattice symmetry axes \cite{Casini07}, which is the case here. 

\begin{figure}[t]
\centerline{\includegraphics[angle=0,width=0.925\columnwidth]{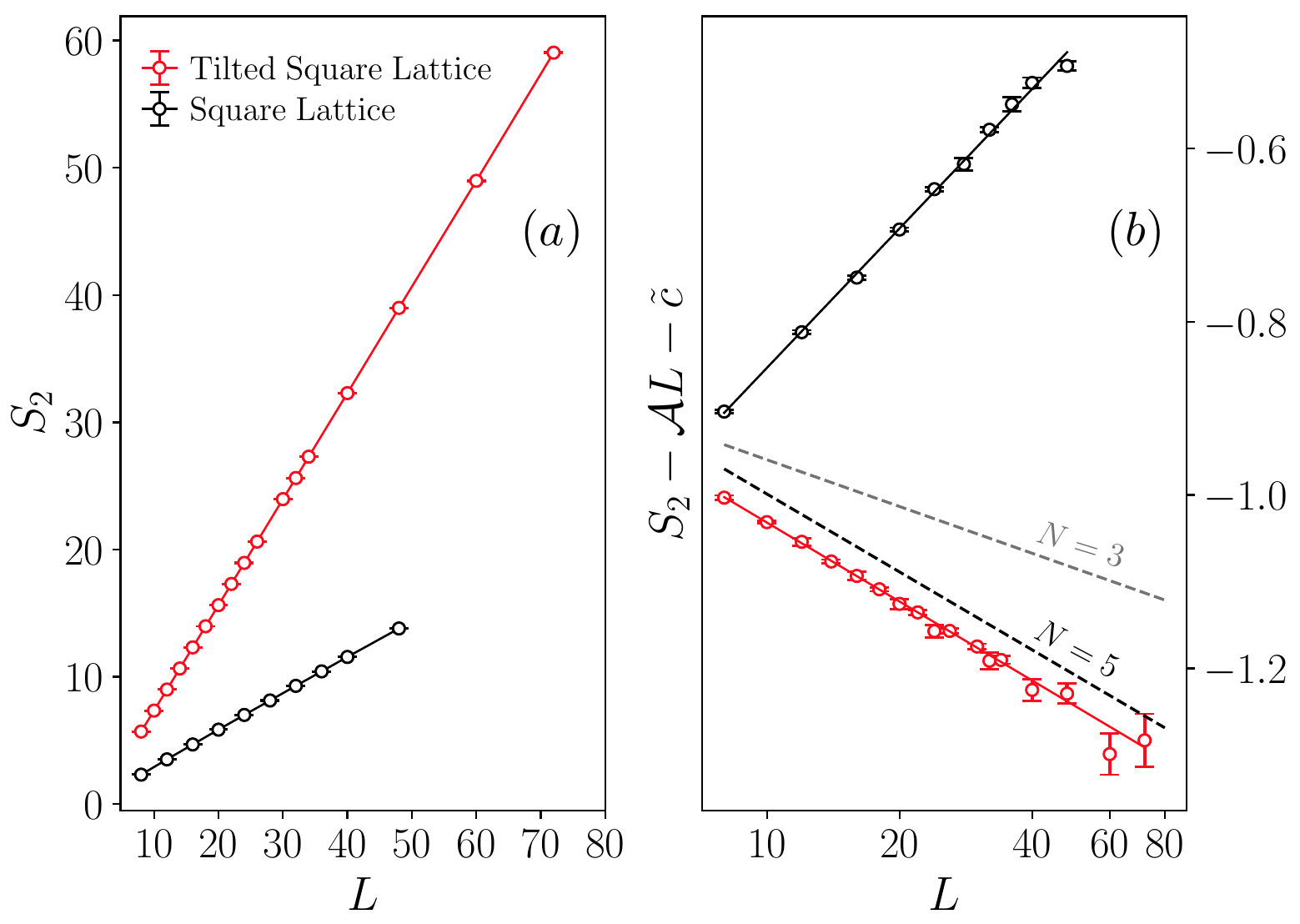}}
\vskip-2mm
\caption{Results for $L\times L$ lattices with subsystem size $L/2 \times L/2$. (a) Size dependence for standard and tilted cuts, along with fits to 
Eq.~(\ref{sform}). (b) The same results with the fitted $\mathcal{A}L$ and constant subtracted off; the square lattice data are shifted by $\tilde{c}=1$ 
to facilitate comparison.  The fit shown gives $a=-0.23(1)$ for the square lattice with a downward drift when dropping small system sizes. For the tilted lattice 
$a=0.131(5)$ with no significant size drift. Predicted slopes for the large-$N$ O($N$) CFTs \cite{Whitsitt17} are shown for $N=3$ and $N=5$. The full fits shown here are $S_2=0.8379(3)L-0.131(5)\log(L)-0.73(1)$ and $S_2=0.2772(5)L+0.23(1)\log(L)-0.39(2)$ for the tilted square lattice and square lattice, respectively.}  
\label{fig:f3}
\end{figure}

%jqEE_tiltedsquareBL
%0.837930866057418
%-0.13147223683418222
%-0.728983847418268
%[0.00028922 0.0052593  0.00969566]
%[ 8. 10. 12. 14. 16. 18. 20. 22. 24. 26. 30. 32. 34. 40. 48. 60. 72.]
%jqEE_sqlattsquareBL
%0.2771828712008129
%0.23217540532476316
%-0.38792246634109007
%[0.00057953 0.01079417 0.0199944 ]
%[ 8. 12. 16. 20. 24. 28. 32. 36. 40. 48.]

Moving on to the cuts with corners, Fig.~\ref{fig:f3} shows results for $L\times L$ lattices of both types. In Fig.~\ref{fig:f3}(a), the SSE data and
three-parameter fit are shown versus $L$ on a linear scale, and in Fig.~\ref{fig:f3}(b) we again use a logarithmic $L$ scale for results with the
area law subtracted away. There are large logarithmic contributions left in both cases, but with the standard cut the coefficient 
has the wrong sign as compared to the CFT expectation, as noted previously \cite{Zhao22,Liao23,Song23a}. This behavior is expected, given the logarithms 
seen for the smooth cut in Fig.~\ref{fig:f2}(b), where the slope for $r=1$ is very close to that in Fig.~\ref{fig:f3}(b).  We also note that the R\'enyi EE of
resonating valence bond wavefunctions was recently shown to contain a similar logarithmic term with the wrong sign \cite{Torlai24}.

For the tilted  lattice, the corner coefficient in Fig.~\ref{fig:f3}(b) has the opposite sign, i.e., in accordance with the CFT. Remarkably, its magnitude 
is very close to the value found for a large-$N$ Wilson-Fisher O($N$) CFT critical point in 2+1 dimensions \cite{Whitsitt17}, which equals that in a free 
(Gaussian) theory with $N$ components. In Fig.~\ref{fig:f3}(b), we have drawn a line corresponding to $N=5$; specifically, the predicted coefficient is 
$a \approx 0.0064875\times4\times5 = 0.12975$, where $0.0064875$ is the value for a free scalar field at a single $90^{\circ}$ corner 
\cite{Helmes16,Whitsitt17}, and we have four corners and five components. A fit to our SSE data gives $a=0.131(5)$, matching the 
free theory result within statistical errors.  While precise agreement with the O($N$) theory is not expected, this large corner contribution strongly suggests an enlarged symmetry at the DQCP.

An SO(5) DQCP has been discussed as a possibility for the AFM--VBS transition \cite{Nahum15b,Wang17,Takahashi20}, and the distinction between O($5$) and SO($5$) 
is not expected to be significant here. The form of the $1/N$ correction is not known, but in the case of the O($3$) transition the value from numerical linked 
cluster simulations \cite{Kallin14} differ by about 15\% from the large-$N$ form with $N=3$---we also show the O($3$) prediction in Fig.~\ref{fig:f3}(b) to 
illustrate the significant $N$ dependence of the predicted form. In \cite{Supp} we explore the corner coefficient for different angles $\theta \neq 90^{\circ}$ where the qualitative
behavior follows the CFT form but with deviations from the large-$N$ form for small angles.

{\em Discussion.}---Taken at face value, the continuum CFT predictions for the EE corner contributions \cite{Casini07} do not depend on the microscopic details 
of how the subsystem is cut out, only on the macroscopic angles. While we have no formal analytical support for qualitative differences between the conventional 
square and tilted cuts of the lattice, our numerical results clearly show very dramatic effects (similar behavior was also observed recently in a fermionic model \cite{Demidio23b}). We have proposed that the AFM--VBS transition is special 
in this regard, because of the VBS dimer fluctuations that are not necessarily accommodated properly by all cuts. This notion guided us in exploring the tilted 
cut, the boundary of which treats equally not only the four columnar VBS patterns but also the four plaquette patterns, whose critical fluctuations in the scenario of emergent 
U($1$) symmetry must be on par with those of the columnar ones \cite{Levin04}. The standard square cuts suppress some of these fluctuations in an effective EE 
Hamiltonian of the subsystem, thereby, apparently \cite{Deng23a}, making visible the Goldstone contributions of the near-critical $J$-$Q$ model. A slightly
different $J$-$Q_3$ model with stronger first-order behavior was used in Ref.~\cite{Deng23a}, and we find a consistent CFT corner logarithm in this model as well \cite{Supp}.  With the tilted cut, we do not find any statistically significant logarithms in the 
absence of corners, and the corner contributions are fully compatible with an SO(5) CFT \cite{{Whitsitt17}}.  See also \cite{Supp} for a complementary method to extract the corner coefficient, which agrees with the results presented here.

We do not expect the tilted lattice to remove all Goldstone contributions from an ultimately long-range ordered coexistence state. Indeed, in the Heisenberg 
model, which we have also studied, the expected logarithmic form is is observed. Most likely, the tilted cut merely suppresses the finite-size ordering due 
to the compatibility of its boundaries with all AFM and VBS critical fluctuations; in contrast, the standard cut does not fully accommodate the VBS 
fluctuations and thereby may act as to amplify the finite-size ordering.

In Refs.~\onlinecite{Zhao22,Liao23,Song23a,Song23b}, the negative corner coefficient and smooth-cut logarithm found with the standard subsystems formed the basis of various 
speculations of non-CFT nature of the AFM--VBS transition, and the results were also used in an attempt
to explain the absence of a DQCP in high-pressure experiments 
on SrCu$_2$(BO$_3$)$_2$ \cite{Guo23} (while other experiments support a proximate DQCP \cite{Cui23}). In light of our results presented here, the scenario 
of an SO(5) tri-critical point \cite{Chester23} located extremely 
close to the transition in the $J$-$Q$ model (and in many other systems) appears much more likely. In this scenario, scaling governed by the underlying 
CFT is manifested up to very large length scales, below which the ultimately stabilized coexisting weak AFM and VBS long-range orders are largely 
inconsequential. The CFT bootstrap calculations \cite{Chester23} suggest that the tri-critical point is that of a conventional 
unitary CFT, at variance with the proposal of an inaccessible non-unitary CFT \cite{Wang17}. Further work on lattice models will be required to 
positively discriminate between these scenarios.

{\em Acknowledgements.}---We would like to thank Wenan Guo, Michael Levin, Zi Yang Meng, Max Metlitski, and Cenke Xu for discussions.
AWS is supported by the Simons Foundation under Simons Investigator Grant No.~511064. 
The numerical calculations were carried out on the Shared Computing Cluster managed by Boston University's Research Computing Services.

\section{Supplemental material}

\noindent \hyperref[sec:alg]{I. Details on equilibrium algorithm for computing $S_2$}\\

\noindent \hyperref[sec:edge]{II. Edge effects of the $\lambda$ protocol}\\

\noindent \hyperref[sec:Nlam]{III. Dependence on the number of intermediate $\lambda$ values}\\

\noindent \hyperref[sec:efficiency]{IV. Comparing the efficiency of the nonequilibrium, nonequilibrium increment, and equilibrium methods}\\

\noindent \hyperref[sec:qmcvsed]{V. QMC versus ED}\\

\noindent \hyperref[sec:evenodd]{VI. Even-odd effects of the tilted square lattice}\\

\noindent \hyperref[sec:angledep]{VII. Corner logs at the DQCP with angle $\theta \neq 90^{\circ}$}\\

\noindent \hyperref[sec:JQ3and4]{VIII. Corner entanglement in the $J$-$Q_3$ and $J$-$Q_4$ models}\\

\noindent \hyperref[sec:cornersub]{IX. Corner subtraction}\\

\noindent \hyperref[sec:finiteT]{X. Insensitivity of corner coefficient with $\beta/L$ ratio}\\

\section{I. Details on equilibrium algorithm for computing $S_2$}
\label{sec:alg}

Here we outline in more detail the algorithm used for computing $S_2$ in this work.  This is an equilibrium variant of the nonequilibrium approach originally introduced in \cite{Demidio20}.  In fact, this equilibrium variant was recently developed in order to compute $S_2$ in fermionic systems \cite{Demidio23b}, while we use it here for the first time in the context of quantum spin systems.

\begin{figure}[!t]
\centerline{\includegraphics[angle=0,width=1.0\columnwidth]{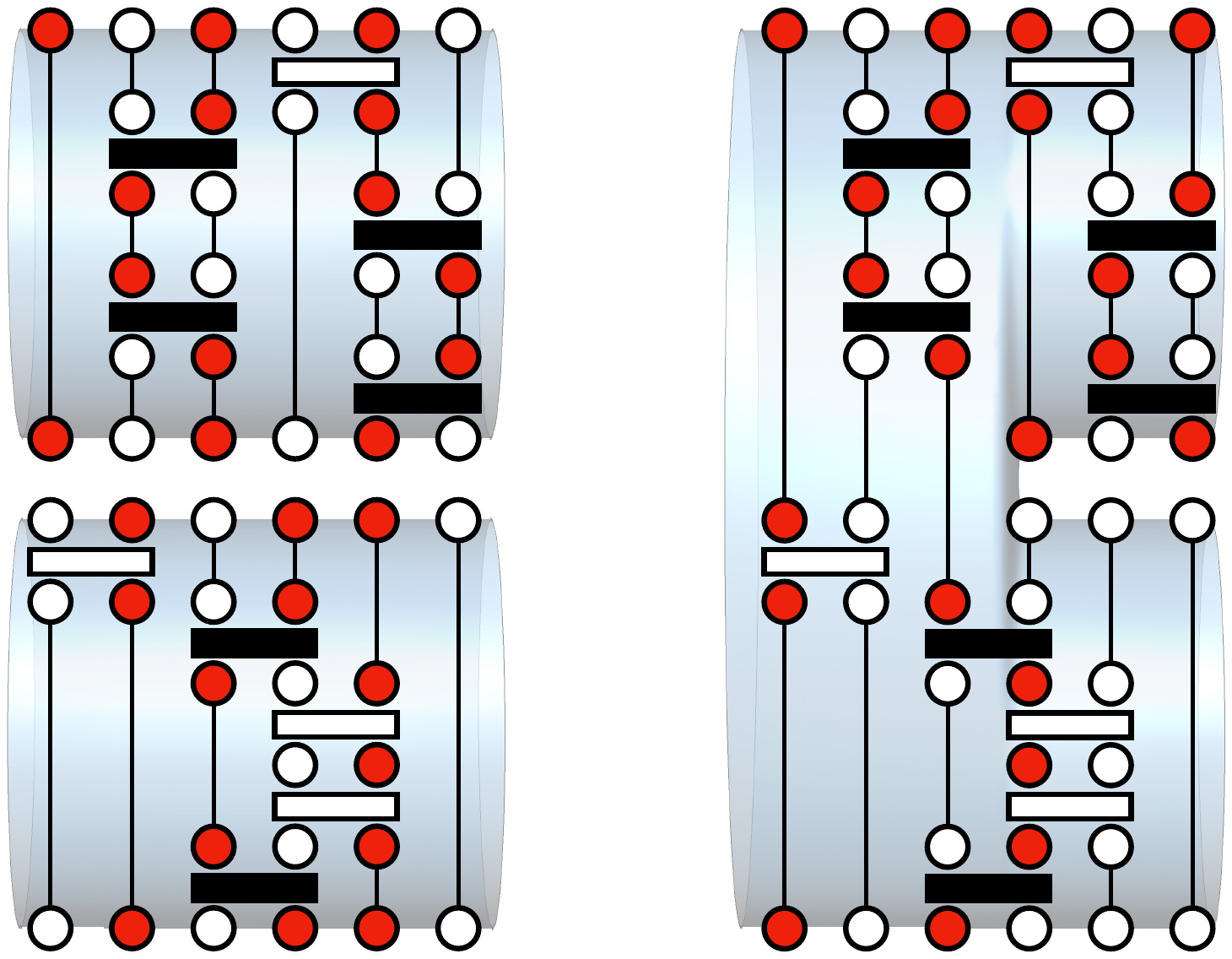}}
\caption{Left: an example SSE configuration \cite{Sandvik10a} of an $L=6$ chain in the $Z_{\varnothing}=Z^2$ ensemble, consisting of two independent traces over the entire system.  Right: an SSE configuration in the $Z_A$ ensemble, consisting of a single trace in the $A$ region (here the left three sites) and two traces in the region $\bar{A}$ (the right three sites).}
\label{fig:configs}
\end{figure}

The starting point for these simulations is the so-called ``replica trick" \cite{Calabrese04}, where one implements the subsystem trace (boundary condition in imaginary time) according to the definition of the R\'enyi entanglement entropies (EE) with integer index: $e^{-(n-1)S_n} = \mathrm{Tr}(\rho^n_A)$, where $\rho_A = \mathrm{Tr}_{\bar{A}}(e^{-\beta H})/Z$ and $Z$ is the partition function.  For $S_2$ we must compute  $\mathrm{Tr}(\rho^2_A) \equiv {Z_A}/{Z_{\varnothing}}$, where $Z_A$ is a replicated partition function where the region $\bar{A}$ is traced twice and $Z_{\varnothing}=Z^2$ (all sites are traced twice), see Fig. \ref{fig:configs}.

The R\'enyi EE can be regarded as the free energy difference between the $Z_{\varnothing}$ and $Z_A$ ensembles.  QMC simulations typically need to be able to move between configurations contained in both ensembles or order to estimate this free energy difference.  In the past, jumping between ensembles was analogous to flipping a switch and changing ensembles whenever the trace condition in the $A$ region would allow it \cite{Humeniuk12}.  The approach that we adopt here is rather akin to introducing a dimmer switch, where one can gradually tune between these two ensembles as a function of a continuous parameter $0<\lambda<1$.

A very efficient interpolating ensemble of this kind was introduced in Ref. \cite{Demidio20}, and is given by
\begin{equation}
\label{eq:Zlam}
\mathcal{Z}(\lambda) = \sum_{C \subseteq A} \lambda^{N_C}(1-\lambda)^{N_A-N_C}Z_C,
\end{equation}
Where the sum over the regions $C$ contains all proper subsets of the region $A$, and $N_C$ and $N_A$ are the number of sites in $C$ and $A$, respectively.  $Z_C$ is a replica partition function with $C$ traced once and $\bar{C}$ traced twice (for the second R\'enyi entropy).

The motivation for this ensemble is to consider each site of the $A$ subsystem individually, and allow it to be traced either once with a weight factor $\lambda$ or traced twice with a weight factor $1-\lambda$.  This way tuning from $\lambda=0$ to $1$ continuously moves from the $Z_{\varnothing}$ ensemble to the $Z_A$ ensemble, and each site in the $A$ region may stochastically change its trace condition independent of the other sites.  When one expands this $\lambda$ weighting as a product over all sites in the $A$ region, one obtains Eq.~(\ref{eq:Zlam}).

QMC simulations of the $\mathcal{Z}(\lambda)$ ensemble are similar to standard SSE simulations, except for the extra update that allows the trace condition on each site of the $A$ region to change depending on if the spin state for this site matches between the different replicas.  Since this matching condition is only required at the level of individual sites in the $A$ region it is very often satisfied. The trace condition of each site can then be modified according to the ratio of the $\lambda$ weight factors satisfying detailed balance.

\begin{figure}[!t]
\centerline{\includegraphics[angle=0,width=0.95\columnwidth]{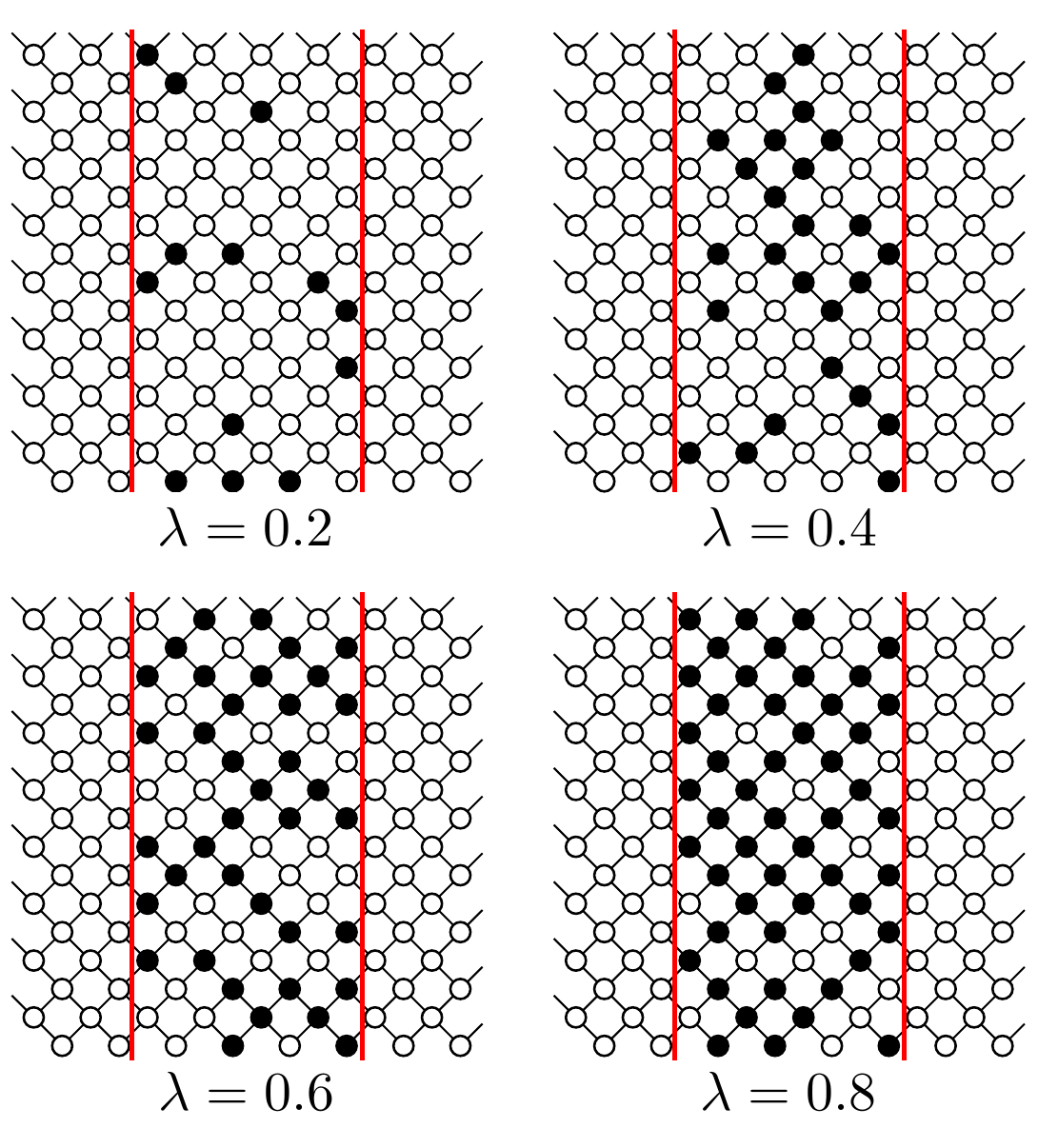}}
\caption{Different configurations of the single-trace sites (region $C$) taken from three actual simulations of an $L=8$ tilted square lattice with $\lambda=0.2,0.4,0.6,0.8$ (15 different $\lambda$'s were used for the actual $S_2$ computation).  The single-trace $C$ sites are colored black and all other sites in white are traced twice.}
\label{fig:lamconfigs}
\end{figure}

The variant of this approach that we use here for the first time with the SSE algorithm is to simulate the $\mathcal{Z}(\lambda)$ ensemble in equilibrium at various $\lambda$ values between 0 and 1.  The only important quantity for these computations is the value of $N_C$, or the number of single-trace sites, see Fig. \ref{fig:lamconfigs}.  Thinking in terms of histograms of the $N_C$ values in each simulation, the different $\lambda$ simulations form a bridge of overlapping histograms of $N_C$ that leads from $N_C=0$ to $N_C=N_A$, allowing for highly efficient computations of the free energy difference between the endpoint ensembles ($Z_{\varnothing}$ and $Z_A$, respectively).

The formula that allows for such computations is the estimator for the partition function ratio at two nearby $\lambda$ values, given by
\begin{equation}
\label{eq:Zrat}
\frac{\mathcal{Z}(\lambda_j)}{\mathcal{Z}(\lambda_i)} = \left\langle \left(\frac{\lambda_j}{\lambda_i}\right)^{N_C}\left(\frac{1-\lambda_j}{1-\lambda_i}\right)^{N_A-N_C} \right\rangle_{\lambda_i}.
\end{equation}
This formula comes from a simple reweighting of Eq.~(\ref{eq:Zlam}) that transforms the weight factors of the $\mathcal{Z}(\lambda_i)$ ensemble to those of the $\mathcal{Z}(\lambda_j)$ ensemble.

An important point is that this formula is only applicable when $0 < \lambda_i < 1$, i.e. the actual simulation must be done at some intermediate values of $\lambda$, and one never simulates the $\mathcal{Z}(0)=Z_{\varnothing}$ or $\mathcal{Z}(1)=Z_{A}$ ensembles.  It is instructive to see how the formula works when computing $S_2$ based on a single intermediate simulation at $\lambda=\lambda_m$ (for instance this could be $\lambda_m=1/2$, but we choose to keep it general here).  The goal is to compute $\frac{Z_A}{Z_{\varnothing}}$. Using our single intermediate state this becomes 
\begin{equation}
\frac{Z_A}{Z_{\varnothing}} = \frac{\mathcal{Z}(1)}{\mathcal{Z}(\lambda_m)} \frac{\mathcal{Z}(\lambda_m)}{\mathcal{Z}(0)}.  
\end{equation}
Note that Eq.~(\ref{eq:Zrat}) can only be used when $\lambda_i$ appearing in the denominator (the one actually being simulated) is neither 0 nor 1.  So we cannot evaluate ${\mathcal{Z}(\lambda_m)}/{\mathcal{Z}(0)}$ and must instead compute its inverse and later invert it.  Focusing first on the computation of ${\mathcal{Z}(1)}/{\mathcal{Z}(\lambda_m)}$, we see that Eq.~(\ref{eq:Zrat}) gives zero if $N_C \neq N_A$, otherwise it gives $1/ \lambda^{N_A}_m$.  Combining this information we have ${\mathcal{Z}(1)}/{\mathcal{Z}(\lambda_m)} = n_{A} / \lambda^{N_A}_m$, where $n_{A}$ is the average number of times the region $C=A$.  In this same way we have 
${\mathcal{Z}(0)}/{\mathcal{Z}(\lambda_m)} =   n_{\varnothing}/(1- \lambda_m)^{N_A}$.
Finally, combining this we have 
\begin{equation}
\frac{Z_A}{Z_{\varnothing}}=\left(\frac{1-\lambda_m}{\lambda_m}\right)^{N_A} \frac{n_{A} }{n_{\varnothing}},
\end{equation}
which simplifies to ${n_A}/{n_{\varnothing}}$ when $\lambda_m=1/2$.

Readers familiar with QMC computations of the R\'enyi EE will recognize the previous formula ${Z_A}/{Z_{\varnothing}} = {n_A}/{n_{\varnothing}}$, which is the same as in the popular approach put forward in Ref. \cite{Humeniuk12}, but here appearing in a different context.  In Ref. \cite{Humeniuk12} transitions are made between the $Z_A$ and $Z_{\varnothing}$ ensemble whenever the trace condition in the entire $A$ region matches between the R\'enyi replicas (when it is possible to join or split the traces).  This becomes exponentially improbable with the size of the region $N_A$, so one must introduce intermediate ensembles where the region $A$ is incrementally built up \cite{Hastings10}.  Similarly here, a single simulation (say at $\lambda_m=1/2$) is exponentially unlikely to find itself in the configuration $C=A$, since this probability for an uncorrelated ensemble is given by $\lambda^{N_A}_m$.  Similarly the probability for $C=\varnothing$ is given by $(1-\lambda_m)^{N_A}$.  So, as in previous approaches, we must introduce more intermediate states, which is conceptually clear given that our single simulation starting point begins already with an intermediate state at $0<\lambda_m<1$.

We end this section by noting that it is advantageous to not directly compute the estimator in Eq.~(\ref{eq:Zrat}) directly during the QMC simulations, as this requires knowledge of the different $\lambda$ values in advance.  Instead, we store a histogram of the values of $N_C$ that are encountered throughout the course of each of the simulations.  This allows Eq.~(\ref{eq:Zrat}) to be applied in post-processing for arbitrary values of $\lambda_j$ in the numerator ensemble.  This allows for the grid of $\lambda$ values to be later refined by adding more simulations if one desires.

\begin{figure}[!t]
\centerline{\includegraphics[angle=0,width=1.0\columnwidth]{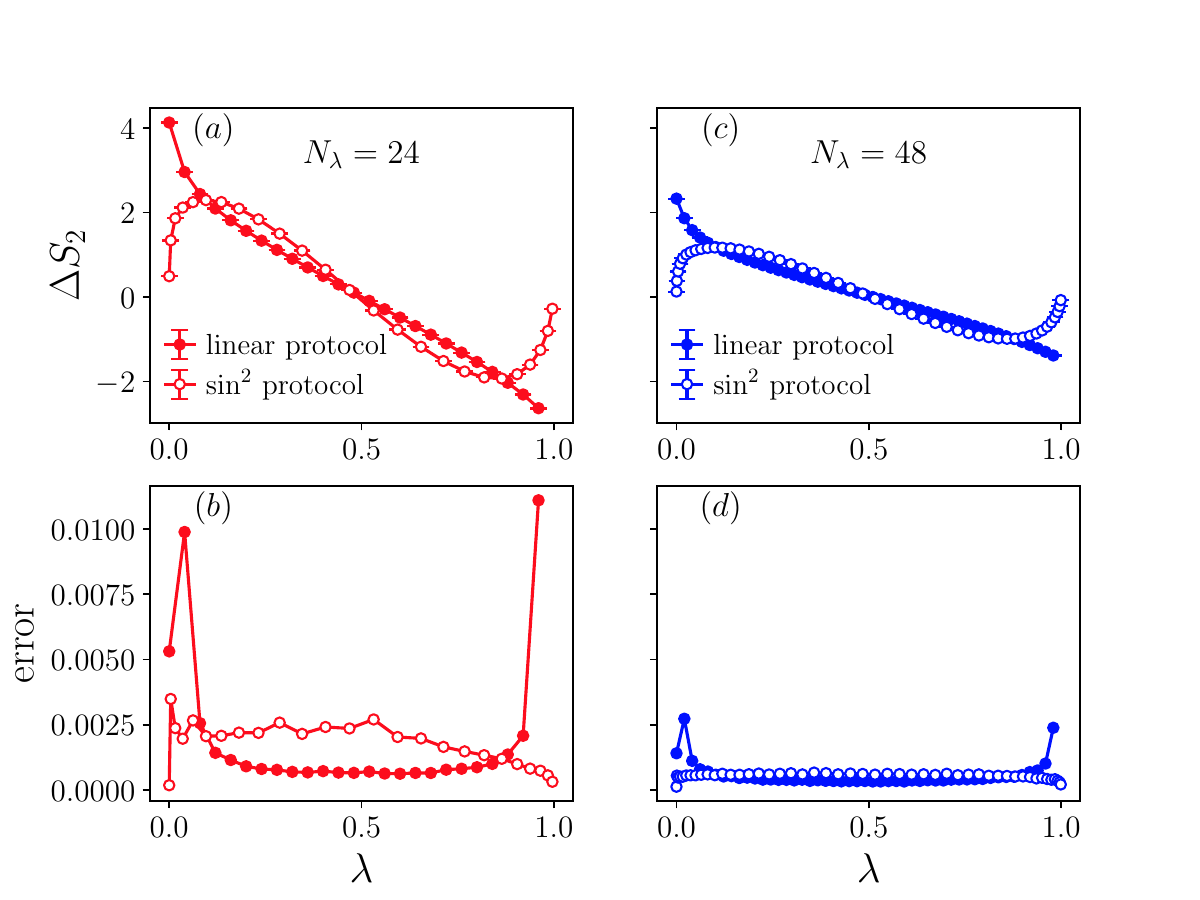}}
\caption{Comparing the stochastic error generated when using a linear versus a $\sin^2$ $\lambda$ discretization.  The linear protocol shows noticeable spikes in the QMC error bar coming from the endpoints.}
\label{fig:qpcomp}
\end{figure}

\section{II. Edge effects of the $\lambda$ protocol}
\label{sec:edge}

We have seen that one must introduce many intermediate states (with different $\lambda$ values between 0 and 1) and use Eq.~(\ref{eq:Zrat}) to compute each factor in the ratio 
\begin{equation}
\frac{Z_A}{Z_{\varnothing}} = \frac{\mathcal{Z}(\lambda_1)}{\mathcal{Z}(0)} \frac{\mathcal{Z}(\lambda_2)}{\mathcal{Z}(\lambda_1)} \ldots \frac{\mathcal{Z}(1)}{\mathcal{Z}(\lambda_N)}.  
\end{equation}
Here we will take a closer look at the stochastic error generated in the computation of each factor individually, and how this depends on the way the $\lambda$ values are spaced (the protocol).

We find that when using a linear grid of $\lambda$ values, $\lambda_i=i/(N_{\lambda}+1)$ with $i=1,..,N_{\lambda}$, the endpoint ratios tend to generate much more stochastic error than the interior ratios.  In order to avoid this, we instead use a $\lambda$ grid that is smooth near the endpoints, given by 
\begin{equation}
\lambda_i=\sin^2\left (\frac{i\pi/2}{N_{\lambda}+1}\right ),~~i=1,..,N_{\lambda}.  
\end{equation}
The comparison of these two protocols is given in Fig. \ref{fig:qpcomp}.  The upper panels show the entropy increments associated with each factor (the sum of which give the total entropy) for and$L_x=32, L_y=16$ square lattice Heisenberg model.  The points in red (left panels) use a discretization of $N_{\lambda}=24$ and the blue points (right panels) use a discretization of $N_{\lambda}=48$.  The bottom panels show the QMC error bar generated from each of the segments.  We observe a notable spike near $\lambda=0,1$ with the linear protocol, whereas $\sin^2$ gives errors that taper off near the endpoints.  In the end the total error of $S_2$ is usually comparable between these two protocols when $N_{\lambda}$ is large, but for smaller discritizations $\sin^2$ gives better results.  Throughout this work we have used the $\sin^2$ protocol.

\section{III. Dependence on the number of intermediate $\lambda$ values}
\label{sec:Nlam}

\begin{figure}[!t]
\centerline{\includegraphics[angle=0,width=1.0\columnwidth]{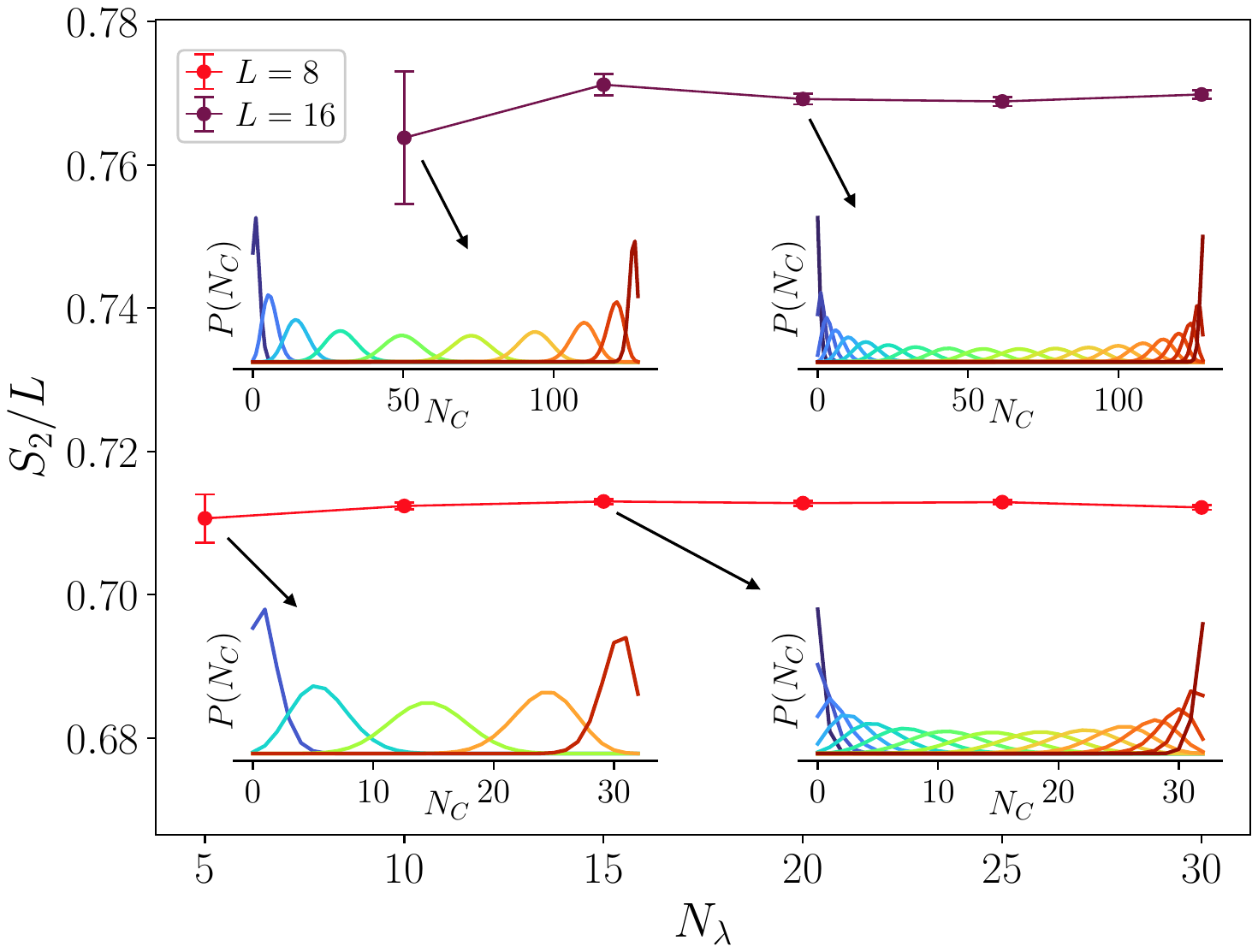}}
\caption{Comparing independent computations of $S_2$ using different numbers of intermediate $\lambda$ values ($N_{\lambda}$).  The histograms are values of $N_C$ encountered during the $N_{\lambda}$ different simulations that go into each data point.}
\label{fig:hist}
\end{figure}

Here we investigate in further detail the dependence on the number of intermediate $\lambda$ values ($N_{\lambda}$) used in the equilibrium method.  To show this we take tilted lattice systems at the DQCP ($J=0.04502$, $Q=1$ and $\beta=L$) and compute the half-system $S_2$.  This is done independently as a function of the number of $\lambda$ values $N_{\lambda}$ ($\sin^2$ protocol).  We also fix the total number of measurement sweeps to be the same between all of these computations.  This is shown in Fig. \ref{fig:hist}, where we show $S_2/L$ as a function of $N_{\lambda}$ for $L=8,16$.  We notice that the QMC error bar is initially large when $N_{\lambda}$ is small, but for larger $N_{\lambda}$ the error bars essentially remain constant.  Thus for large enough $N_{\lambda}$ adding more $\lambda$ values does not change the efficiency. This coincides with the overlap of the histograms of $N_C$ between neighboring $\lambda$ values.  These histograms are depicted for four different data points indicated with arrows.  We see that a large error bar is generated when the histograms have relatively small overlap.  Once the histograms are reasonably well overlapped, the error bars are essentially converged.  We note that the sufficient $N_{\lambda}$ for converged error bars is small compared the the values of $N_{\lambda}$ that we have used in this work.  Typically we take $N_{\lambda} \propto L^{\alpha} $ with $\alpha \in [1.2,1.4]$.

\section{IV. Comparing the efficiency of the nonequlibrium, nonequilibrium increment, and equilibrium methods}
\label{sec:efficiency}

In this section we make a direct comparison of the various methods for computing $S_2$ that make use of the $\mathcal{Z}(\lambda)$ ensemble.  In the original nonequilibrium algorithm (Neql) \cite{Demidio20} one performs a nonequilibrium quench by varying $\lambda$ from 0 to 1 in a single QMC simulation and computes the work performed in the process.  Jarzynski's equality (using the average of the exponential of the work) gives a statistically exact estimator for the R\'enyi entanglement entropy.  A variant of this approach is to break the quench up into separate chunks (increments) and to compute these separately \cite{Zhao22}.  We refer to this as the nonequilibrium increment (Neql-Inc) approach.  Finally we have the equilibrium method (Eql) that we use in this work.  

\begin{figure}[!t]
\centerline{\includegraphics[angle=0,width=1.0\columnwidth]{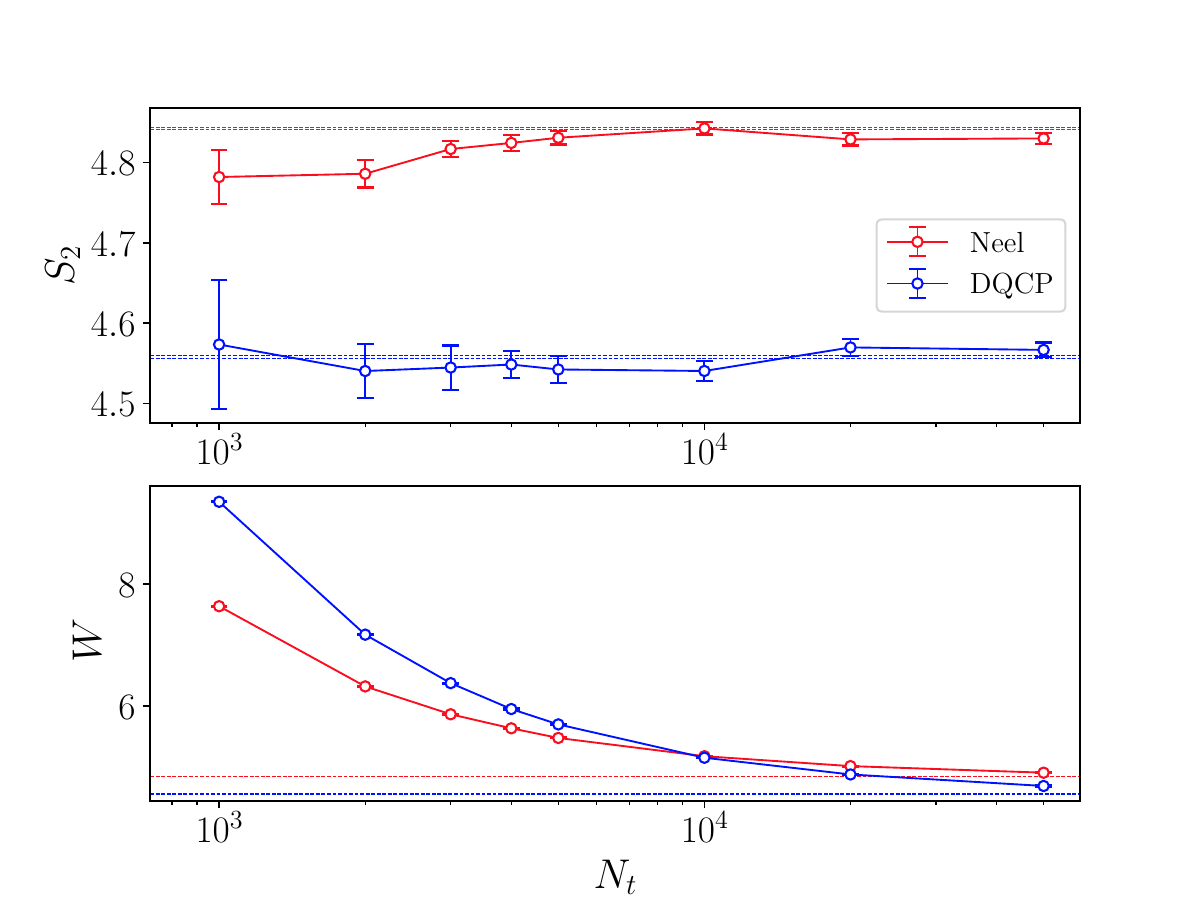}}
\caption{The convergence of the original nonequilibrium (Neql) algorithm as a function of the quench time, in both the N\'eel phase and at the DQCP.  Here we find fast convergence of the Jarzyski estimator for $S_2$ with a potential bias for very small $N_t$.  Even though unbiased results can be obtained for small $N_t$, as shown here, it is typically best to work in the limit where $N_t$ is much larger.  In this limit the Jarzynski estimator and the average work (bottom panel) give the same value.}
\label{fig:neqlconverg}
\end{figure}

Before comparing these three methods directly, we first would like to demonstrate the performance of the original Neql approach, and see how it depends on the quench time $N_t$.  We will show this for the N\'eel state ($Q=0$) as well as at the DQCP.  In Fig. \ref{fig:neqlconverg} we show $S_2$ obtained with the original Neql approach for an $L_x=32,L_y=16$ square lattice system ($\beta=32$) as a function of the quench time $N_t$ in the N\'eel phase ($J=1$, $Q=0$) and at the DQCP ($J=0.04502$, $Q=1$).  We used 1000 re-equilibration sweeps of our initial starting configurations before each nonequilibrium realization.  In each case the total number of measurement sweeps ($N_t \times N_{bin}$) is held fixed, so the same amount of CPU measurement time is spent for each data point.  At least for the N\'eel phase, we only see a potential slight bias when $N_t$ is made small ($N_t<10^4$).  Typically this method works best when $N_t$ is made much larger and is essentially in the quasi-static regime.  In this regime, the average of the work and the Jarzynski estimator give essentially the same value.  Nevertheless, even for much shorter quench times, as we show here, we find that accurate results can be obtained.  As a comparison, in the lower panel of Fig. \ref{fig:neqlconverg} we show the average work, which massively overestimates the entropy for these short quench times.

\begin{figure}[!t]
\centerline{\includegraphics[angle=0,width=1.0\columnwidth]{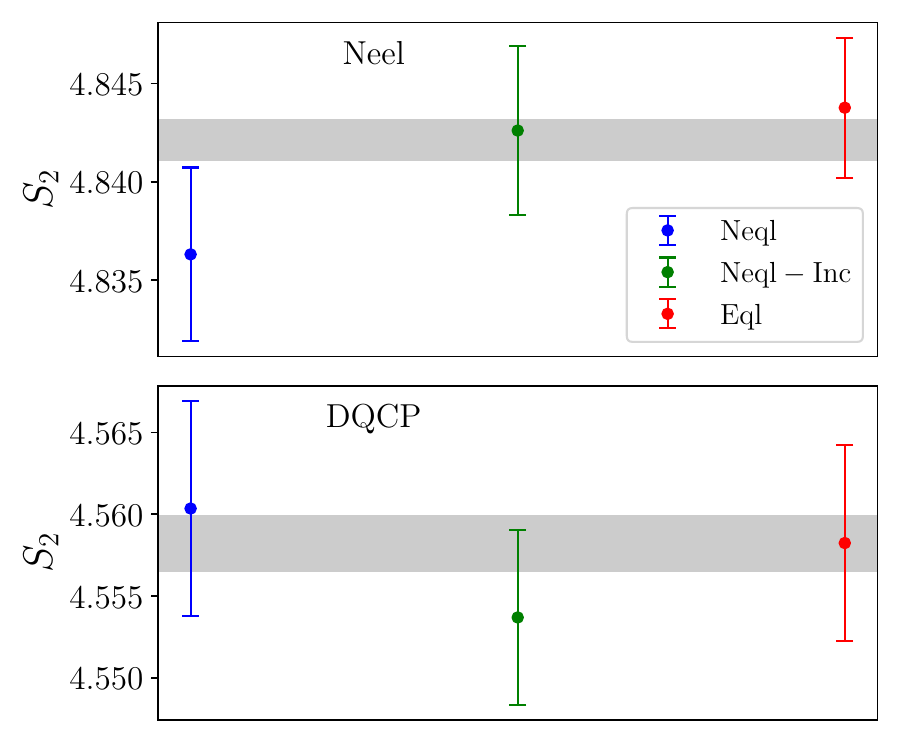}}
\caption{A comparison of three different variants of $S_2$ computations using the $\mathcal{Z}(\lambda)$ ensemble  in the N\'eel phase ($J=1$ $Q=0$) and at the DQCP ($J=0.04502$ $Q=1$) for the half-entropy of an $L_x=32, L_y=16$ square lattice system with $\beta=32$.  We observe that all methods are equally efficient, as shown by comparable error bars given an equal number of measurement sweeps.  We note that the original Neql method with a short ($N_t=10^4$) nonequilibrium quench time performs well compared to the other techniques.}
\label{fig:methodcomp}
\end{figure}

Now we wish to compare all three methods side by side using exactly the same number of measurement sweeps in each case.  This is shown in Fig. \ref{fig:methodcomp}, both in the N\'eel phase and at the DQCP.  Again we focus on an $L_x=32,L_y=16$ system with $\beta=32$ as before.  Here we fix a relatively short quench time of $N_t=10^4$ for the Neql approach, and for the Neql-Inc approach we use $N_t=10^4$ for each of 24 different $\lambda$ windows between 0 and 1 (this gives an effective quench time of $24 \times 10^4$).  For the Eql approach we simulate $N_{\lambda}=48$ different equilibrium $\lambda$ values with the $\sin^2$ protocol.  For each of the methods the number of measurement bins is adjusted so that the exact same number of measurement sweeps is used in each case.  Fig. \ref{fig:methodcomp} shows the computed values of $S_2$ in each case, as compared to grey bars which are the result of more precise runs using the Eql technique.  We conclude that all three methods give essentially the same efficiency (QMC error bar) and are all statistically consistent in value.  We find that any of the methods can fail when either the quench time is made unreasonably short or if equilibrium simulations are spaced too far apart.  However when these parameters are increased, the error bars quickly converge and are consistent between all methods.

\begin{figure}[!t]
\centerline{\includegraphics[angle=0,width=1.0\columnwidth]{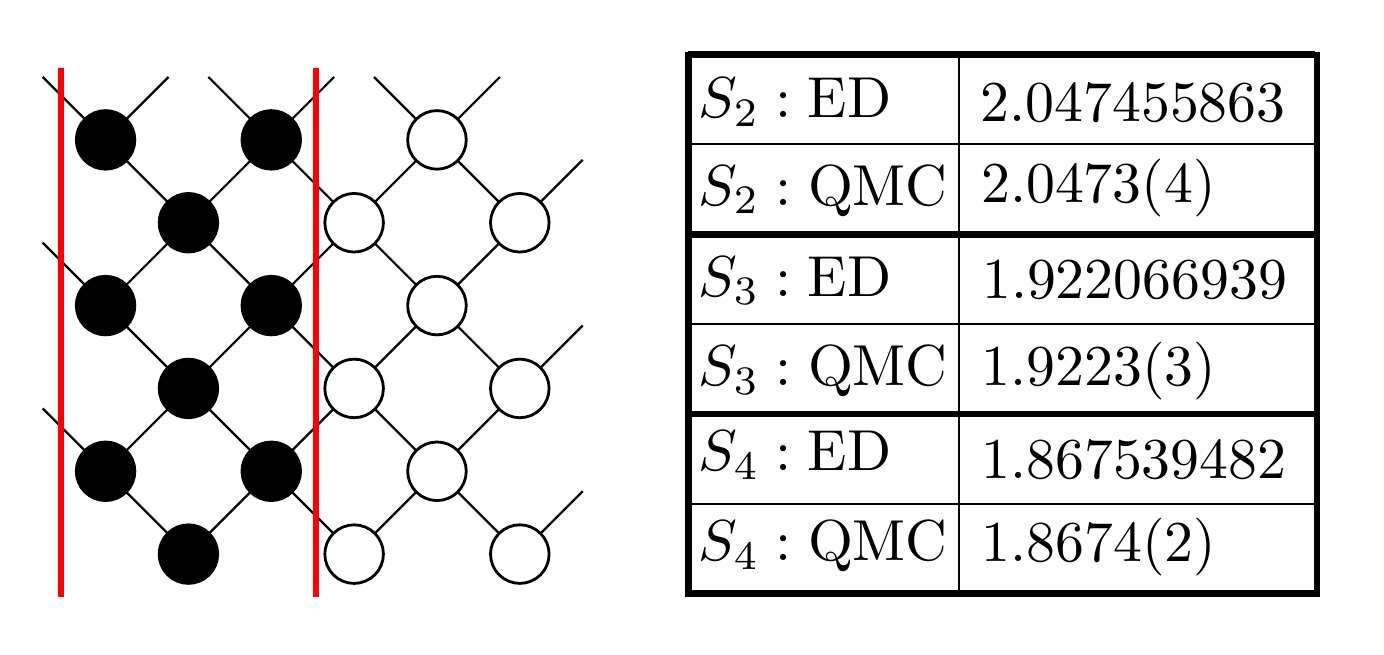}}
\caption{We compute the R\'enyi entanglement entropies ($n=2,3,4$) for the region depicted on an $L=3$ tilted square lattice.  Here we set $J=0.04502$, $Q=1$ and $\beta=40$.}
\label{fig:qmced}
\end{figure}

\section{V. QMC versus ED}
\label{sec:qmcvsed}

Finally we wish to compare our equilibrium QMC method for computing $S_n$ against exact results obtained for a small system.  Here we take and $L=3$ tilted lattice and compute the half system entropy, depicted on the left in Fig. \ref{fig:qmced}.  We have computed $S_2$ with $J=0.04502$, $Q=1$ and for the QMC we take $\beta=40$.  The table shows perfect agreement of the QMC values with exact results for $S_2$ as well as the higher R\'enyi entanglement entropies $S_3$ and $S_4$.

\section{VI. Even-odd effects of the tilted square lattice}
\label{sec:evenodd}

\begin{figure}[!t]
\centerline{\includegraphics[angle=0,width=1.0\columnwidth]{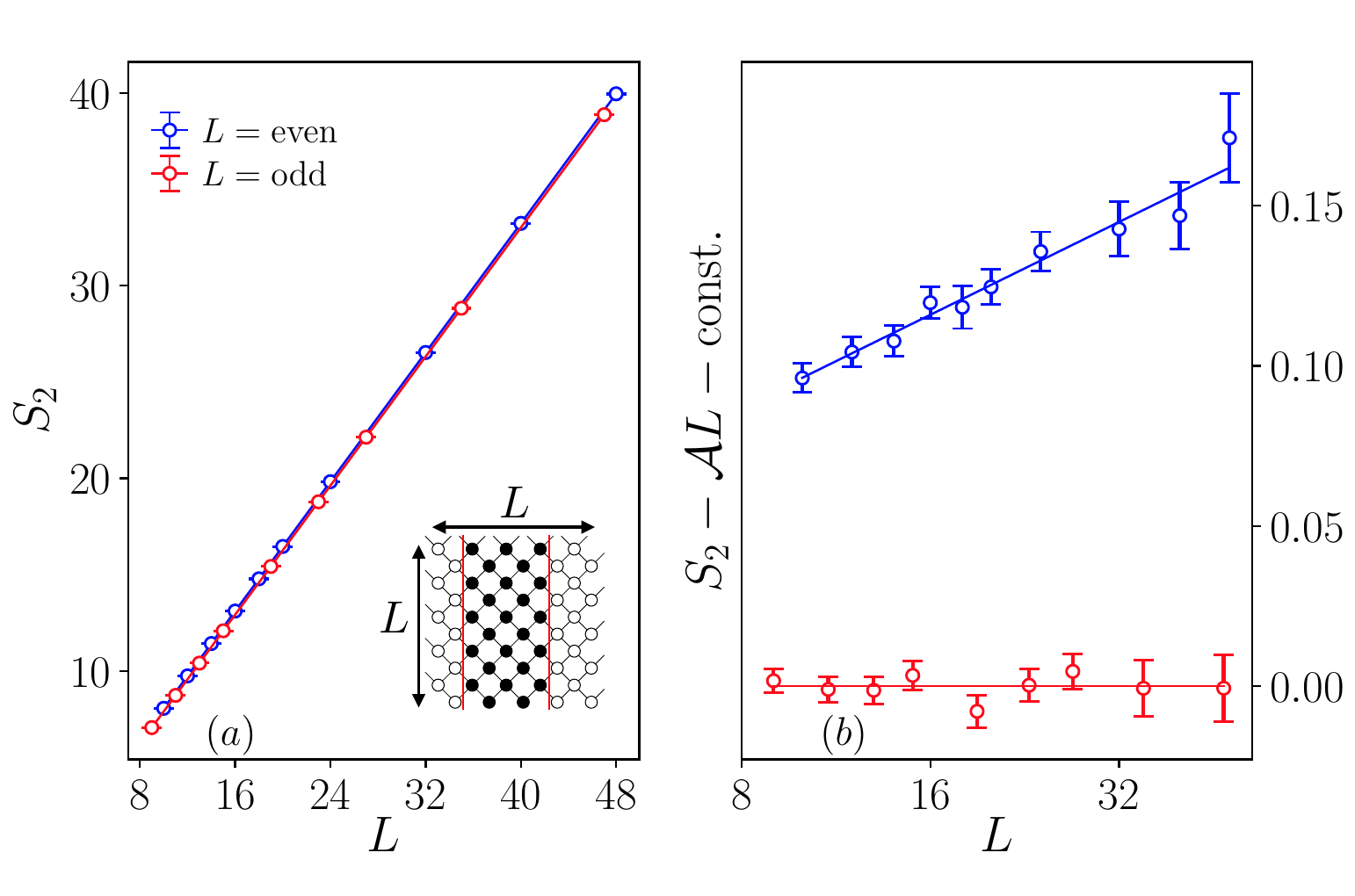}}
\caption{The different finite size scaling behavior of even versus odd size tilted lattices at the DQCP, for a smooth half-cylinder cut.  The even systems seem to produce a small logarithmic contribution, as observed in panel $(b)$ with fitted form $S^{\text{even}}_2=0.8372(5)L+0.04(1)\ln(L)-0.40(2)$.  The odd sizes show a complete absence of any corrections to scaling beyond the area law.  For the odd sizes a linear plus log fit gives $S^{\text{odd}}_2=0.8374(6)L-0.01(1)\ln(L)-0.46(2)$.  For the odd sizes depicted here we have forced a pure area law fit, and panel $(b)$ shows shows the residual of the fit.}
\label{fig:eosmoothR2}
\end{figure}

We now discuss a subtlety that is present in our tilted square lattice systems: the presence of an even/odd system size effect.  Since the tilted lattice is defined with a two site unit cell, it is bipartite with periodic boundaries for both even and odd sizes $L$.  In Fig. \ref{fig:eosmoothR2} we show $S_2$ for a smooth (half cylinder) cut on the tilted lattice at the DQCP, looking at the differences between even sizes versus odd sizes.  We observe a small even/odd oscillation in the raw data and, regarding the two series separately, we see a small positive logarithmic contribution that can be detected here for even sizes.  This is shown in panel $(b)$ by subtracting off the area law piece after a linear plus log three-parameter fit.  We observe that the odd size data seems to show a complete absence of any logarithmic piece.

\begin{figure}[!t]
\centerline{\includegraphics[angle=0,width=1.0\columnwidth]{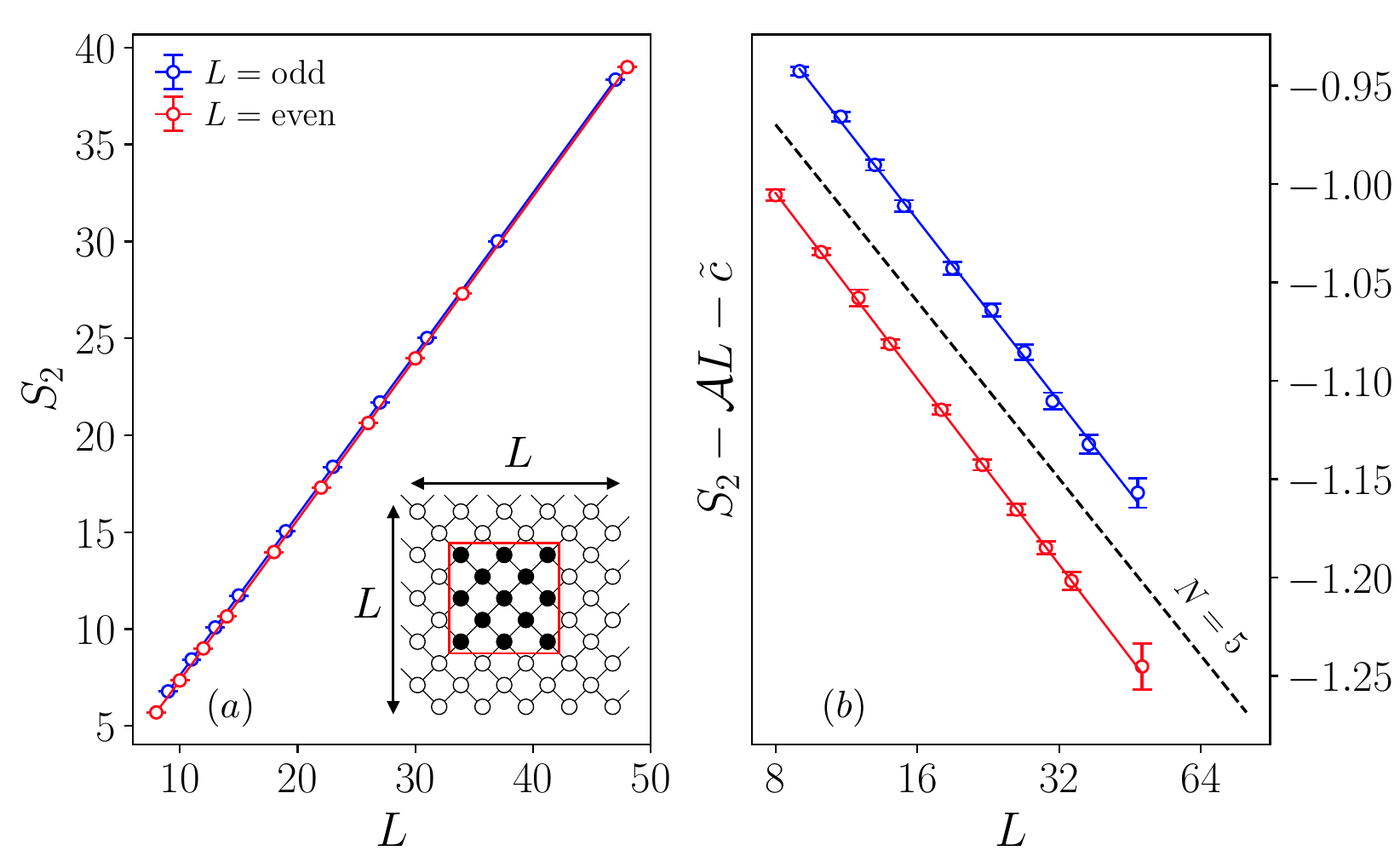}}
\caption{The differences between even and odd systems with regard to the corner coefficient.  A shift of $\tilde{c}=0.2$ is given to the odd sizes for ease of comparison.  Here odd sizes give a corner coefficient $a=0.133(7)$ and even sizes give $a=0.136(3)$ both in agreement with the large-$N$ SO(5) value.  The full fitted forms shown here are $S^{\text{odd}}_2=0.8362(4)L-0.133(7)\ln(L)-0.45(1)$ and $S^{\text{even}}_2=0.8383(2)L-0.136(3)\ln(L)-0.722(5)$.}
\label{fig:eocorner}
\end{figure}

The fact that such differences can arise when studying even versus odd system sizes on the tilted lattice causes concern about the accuracy of the extracted corner coefficient presented in the main text.  Here we compare even versus odd sizes when computing $S_2$ for a square region ($L/2 \times L/2$) on an $L \times L$ tilted square lattice.  This is shown in Fig. \ref{fig:eocorner}, where we depict how the square entangling region is defined for odd sizes in the left panel (using an $L=5$ system as an example).  Again here we find that even/odd oscillations are present in $S_2$, although this is somewhat less surprising in this case given that the entangling region is defined differently for the odd sizes.  We again treat the even and odd series separately and perform three parameter fits to linear plus logarithmic terms.  These extracted log terms are revealed in the right panel, where we see that both series produce nearly identical (critical) logarithmic pieces, both agreeing with the large-$N$ SO(5) value shown by the dashed line.

\section{VII. Corner logs at the DQCP with angle $\theta \neq 90^{\circ}$}
\label{sec:angledep}

By using the tilted square lattice, with a two site unit cell, we naturally make $45^{\circ}$ cuts of the square lattice and make our subsystems a fixed proportion (e.g. $L/2 \times L/2$ unit cells) of the total system size.  We can also generalize this to larger unit cells of sites, arranged along a line, such that $L/2 \times L/2$ subsystems define a parallelogram region where the angles are different from $90^{\circ}$ degrees.  We wish to investigate this here, in order to get a broader sense for the behavior of the corner logarithmic coefficient at the DQCP as a function of the angle.

\begin{figure}[!t]
\centerline{\includegraphics[angle=0,width=1.0\columnwidth]{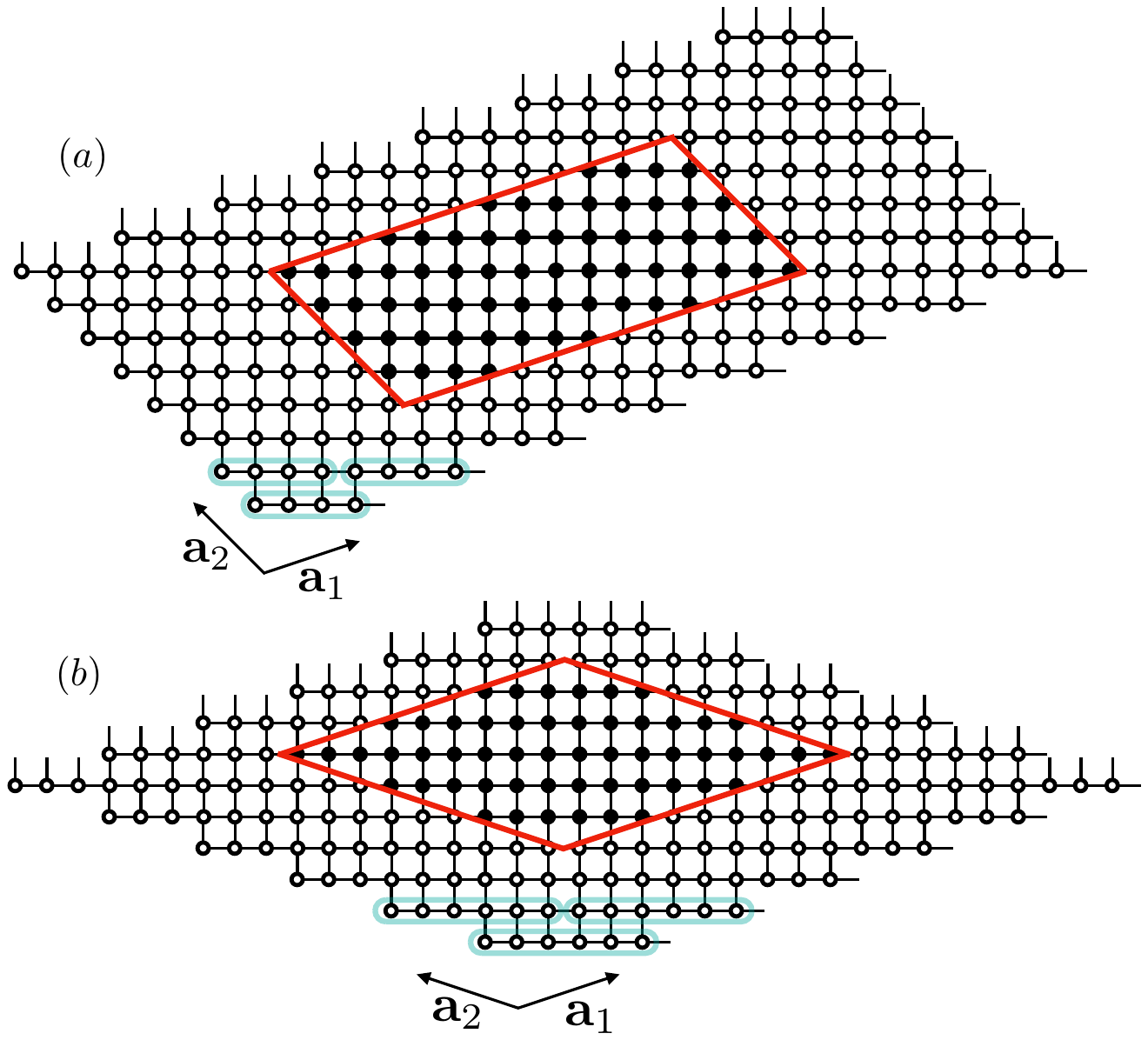}}
\caption{$(a)$: A lattice with a four site unit cell (unit cells shown in light green). Here $L_1=L_2= 8$. $(b)$: A lattice with a six site unit cell ($L_1=L_2=6$).}
\label{fig:tu4and6latt}
\end{figure}

In Fig. \ref{fig:tu4and6latt} we show the lattices that we use here to explore different angles, using a four site unit cell and six site unit cell in panel $(a)$ and $(b)$, respectively.  The need for even numbers of sites in the unit cell, for this definition, is to avoid making cuts that prefer certain VBS patterns over others.

\begin{figure}[!t]
\centerline{\includegraphics[angle=0,width=1.0\columnwidth]{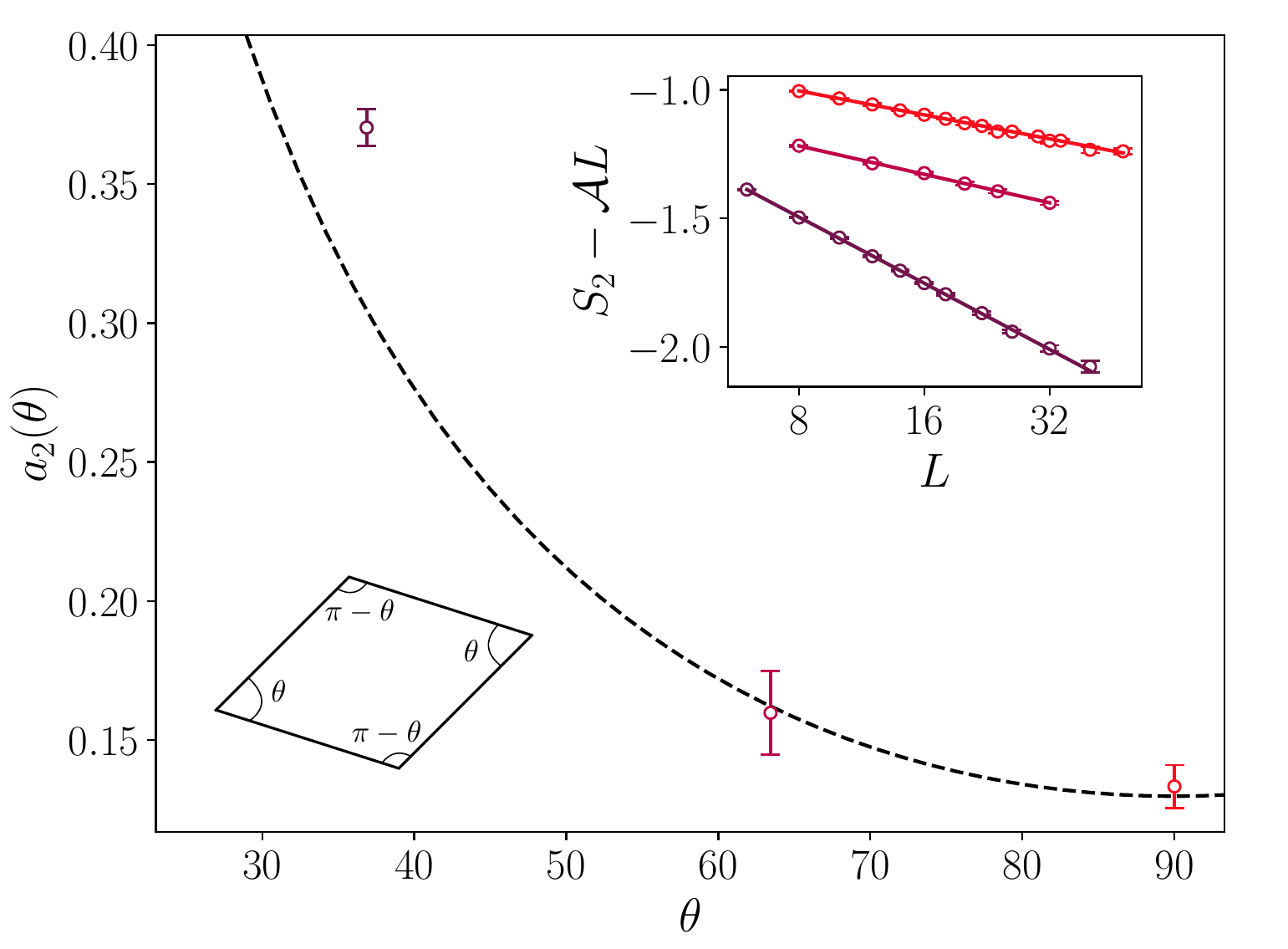}}
\caption{The corner coefficients (main panel) as a function of the acute angle $\theta$, along with the numerical fits to the log term in the inset.  We find that the small angle is significantly larger than the large-$N$ SO(5) form (dashed line), but that the correct overall behavior of the total logarithmic contribution is observed.}
\label{fig:angle}
\end{figure}

In Fig.~\ref{fig:angle} we show the angular dependence of the extracted corner coefficients at the DQCP from the four site and six site unit cell lattices, along with the two site unit cell tilted lattice in the main text.  The inset shows the total logarithmic contribution coming from all corners (two at $\theta$ and two at $\pi - \theta$) after the area law piece is subtracted away, along with the fit.  The numerical values for the coefficients are plotted in the main panel.  The total logarithmic contribution grows in magnitude as the acute angle decreases, with the six site unit cell lattice ($\theta \approx 36^{\circ}$) giving the largest total contribution.  This is compared with the theoretical prediction for a 5 component scalar field \cite{Helmes16} given by the dashed line.

We find that in the small angle case, there are statistically significant deviations away from the large-$N$ SO(5) form.  While it is difficult to know if all lattice effects have been controlled for, we believe this to be a significant result.  Indeed, it is not expected that the corner term should match perfectly the $N=5$ form.  Importantly, we do find the correct dependence in terms of an increasing corner coefficient from small angles.
\\

\begin{figure}[!t]
\centerline{\includegraphics[angle=0,width=1.0\columnwidth]{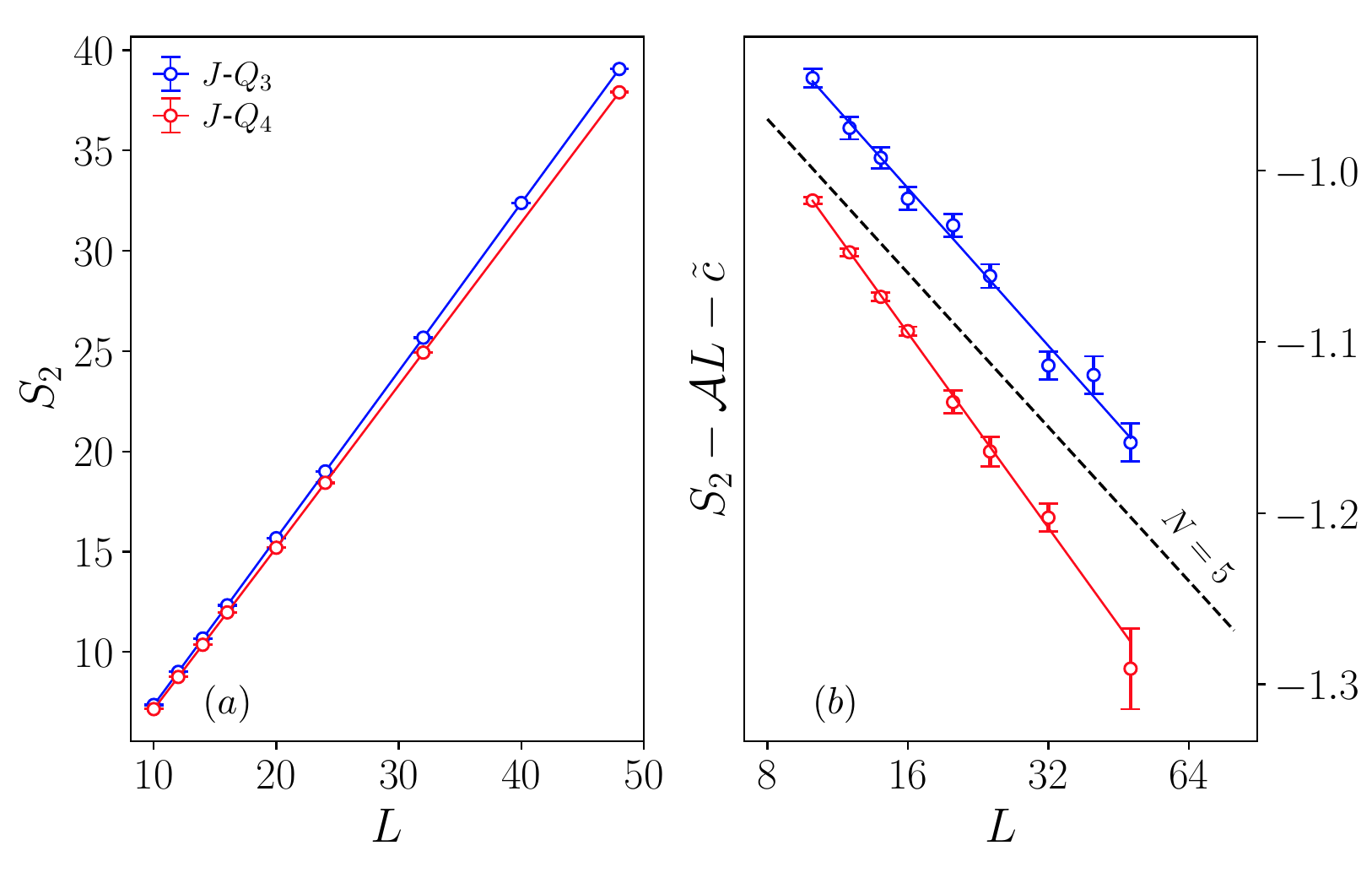}}
\caption{The R\'enyi EE scaling of square subsystems at the deconfined quantum critical point in the $J$-$Q_3$ and $J$-$Q_4$ models using tilted lattices.  For $L_{\text{min}}=10$ to $L_{\text{max}}=48$ pictured here, the fit for the $J$-$Q_3$ model gives $S^{J\text{-}Q_3}_2 = 0.840(1)L - 0.13(2)\ln(L)-0.74(5)$ and the $J$-$Q_4$ model gives $S^{J\text{-}Q_4}_2 = 0.8165(6)L - 0.16(1)\ln(L)-0.64(2)$.}
\label{fig:jq3jq4corner}
\end{figure}

\section{VIII. Corner entanglement in the $J$-$Q_3$ and $J$-$Q_4$ models}
\label{sec:JQ3and4}

Here we compute the corner contribution to the R\'enyi EE in both the $J$-$Q_3$ and $J$-$Q_4$ models.  For the $J$-$Q_3$ model we use the transition point $J=0.67046$ and $Q_3=1$ ($\beta = L$) and for for the $J$-$Q_4$ model we use $J=1$ and $Q_4=0.7024$ and ($\beta=L$), which corresponds to $Q_4=0.4126$ in units where $J+Q_4=1$ \cite{Takahashi23}.  Figure \ref{fig:jq3jq4corner} shows the R\'enyi EE scaling in both models up to $L=48$ on tilted square lattices with a square subsystem.  We find agreement between the logarithmic term and the large-N SO(5) field theory prediction in the $J$-$Q_3$ model, whereas in the $J$-$Q_4$ model a deviation seems to appear.  Interestingly, this deviation seems to indicate a corner coefficient that is larger in magnitude (more negative), contrary to what one would naively expect as the transition becomes more strongly first order with a larger ordered moment.  It is also possible that a better estimate for the transition point in the $J$-$Q_4$ model could lower the value of the corner term significantly.

\begin{figure}[!h]
\centerline{\includegraphics[angle=0,width=1.0\columnwidth]{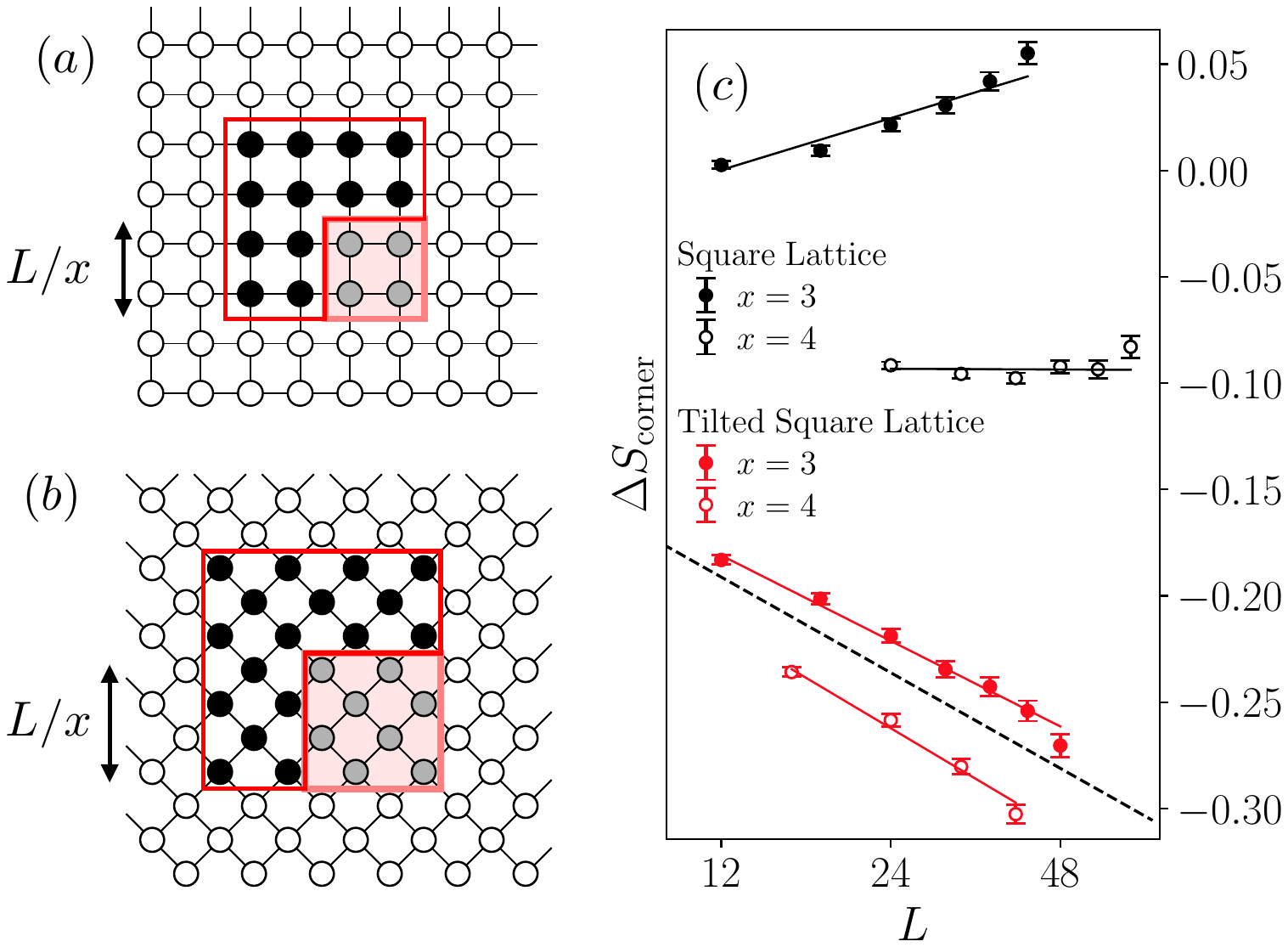}}
\caption{(a) and (b): The corner subtraction scheme used to isolate the logarithmic contribution to $S_2$ on the square lattice and tilted square lattice.  We consider two different cases for both the square and tilted square lattices, one in which the geometry consists of regions with proportion $L/4$ and also $L/3$. (c): $\Delta S_{\mathrm{corner}}$ plotted on a semi-log scale.  We find that the tilted square lattice shows results that are consistent with extracted corner coefficient in the main text.  The dashed line has a slope $-2 \times 5 \times a_{\mathrm{gaussian}} \approx -0.06487$ for comparison. Conversely, the regular square lattice is sensitive to the proportion $L/x$, and the $L/4$ case seems to indicate a nearly zero logarithmic corner contribution.}
\label{fig:cornersubtraction}
\end{figure}
%$x=3$
%0.035132280507613856
%-0.0870496785828441
%[0.00484872 0.01421543]
%[12. 18. 24. 30. 36. 42.]
%$x=4$
%-0.0004918659335328337
%-0.09168969108380884
%[0.0050741  0.01779978]
%[24. 32. 40. 48. 56. 64.]
%$x=3$
%-0.05803435141274928
%-0.036788781960944435
%[0.00293174 0.00894748]
%[12. 18. 24. 30. 36. 42. 48.]
%$x=4$
%-0.06844758874849399
%-0.04476771555926736
%[0.00608486 0.01896326]
%[16. 24. 32. 40.]

\section{IX. Corner subtraction}
\label{sec:cornersub}

Here we explore a complementary approach to compute the corner coefficient at the deconfined critical point on both the tilted and regular square lattice geometries.  The aim is to find consistency between different approaches in order to build confidence in our main results.  To this end, we devise a subtraction scheme in order to directly isolate the logarithmic contribution coming from the corners of the subsystem.  Our setup is depicted in panels (a) and (b) of Fig. \ref{fig:cornersubtraction}.  We compute $S_2$ for the subsystem with the pink region missing, and subtract from that $S_2$ for the subsystem that includes the pink region.  This difference is defined to be $\Delta S_{\mathrm{corner}}$.  As such, the grey sites are the only fluctuating sites in our simulations.  Since these two types of subsystems have the same perimeter but differ by the presence of two extra corners, we expect the scaling $\Delta S_{\mathrm{corner}} = -2a \log(L) + \mathrm{constant}$.  Additionally, we consider two different cases for both the square and tilted square lattices, one in which the geometry consists of regions with proportion $L/x$ with $x=3,4$. 

In panel (c) of Fig. \ref{fig:cornersubtraction} we show our $\Delta S_{\mathrm{corner}}$ data.  We find that the tilted square lattice gives consistent results with the analysis performed in the main paper, namely the direct fit to area-law plus log for a square subsystem.  The obtained coefficient for the logarithmic term is close to $2 \times 5 \times a_{\mathrm{gaussian}} \approx 0.06487$, yielding $\Delta S_{\mathrm{corner}} = -0.058(3)\log(L) - 0.037(9)$ and $\Delta S_{\mathrm{corner}} = -0.068(6)\log(L) - 0.05(2)$ for $x=3,4$, respectively.  We note that a possible downward drift seems to occur at larger sizes, which could arise due to weakly first-order behavior.

Turning to the the regular square lattice data in Fig. \ref{fig:cornersubtraction}, we find that the results are sensitive to the proportion $L/x$ that is used.  Interestingly, for $x=4$ the data seems nearly consistent with zero corner coefficient, albeit with a possible upturn at the largest system size.  For $x=3$ we find a positive slope again with a possible upturn that seems to deviate from a clean $\log(L)$ scaling form.  The fit parameters here are $\Delta S_{\mathrm{corner}} = 0.035(5)\log(L) - 0.09(1)$ and  $\Delta S_{\mathrm{corner}} = 0.000(5)\log(L) - 0.09(2)$ for $x=3,4$, respectively, but with a possible drift as can be seen.

\begin{figure}[!h]
\centerline{\includegraphics[angle=0,width=1.0\columnwidth]{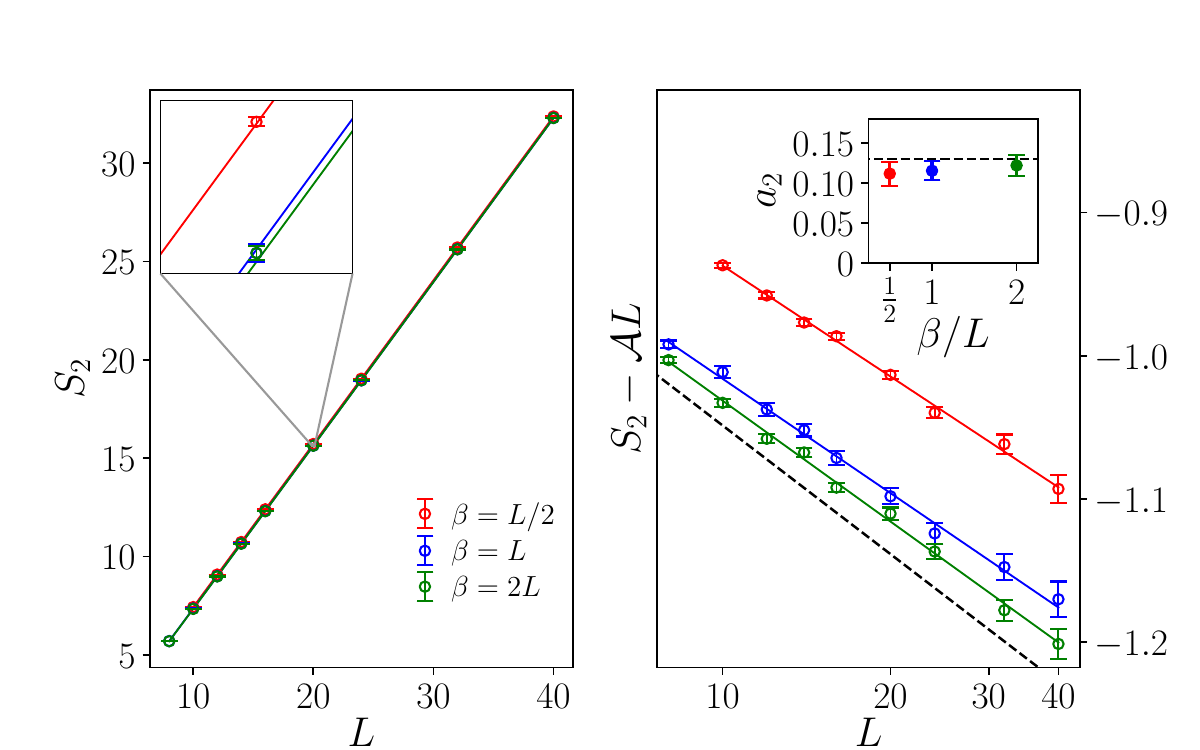}}
\caption{$S_2$ data for the $J$-$Q$ model on the tilted square lattice at the critical point ($J=0.04502$, $Q=1$) for different inverse temperature scalings $\beta/L=0.5,1,2$.  On the right we perform a three parameter fit to the area law plus log plus constant, and subtract away the area law and plot on a logarithmic scale in $L$.  We find the resulting fits to the logarithmic term to be consistent between data sets.  The dashed line indicates the 5-component Gaussian value for comparison.  Note that the raw $S_2$ data is essentially converged to $T=0$ for $\beta/L=1,2$ as shown by the zoom-in in the left panel.  The fits are $S_2 =0.8365(9)L - 0.11(2)\log(L) - 0.68(3), S_2 =0.8366(7)L - 0.12(1)\log(L) - 0.75(2), S_2 =0.8372(8)L - 0.12(1)\log(L) - 0.75(2)$ for $\beta/L = 0.5, 1, 2$, respectively.}
\label{fig:cornerbeta}
\end{figure}
%0.8364895550067206
%-0.11173163102817951
%-0.6798370746755237
%[0.00086114 0.01498459 0.02732489]
%[10. 12. 14. 16. 20. 24. 32. 40.]
%0.8365(9)L - 0.11(2)\log(L) - 0.68(3)
%
%0.8365654227771937
%-0.11533048631039199
%-0.7505544573457422
%[0.00073881 0.01189046 0.02033449]
%[ 8. 10. 12. 14. 16. 20. 24. 32. 40.]
%0.8366(7)L - 0.12(1)\log(L) - 0.75(2)
%
%0.8371692412658818
%-0.12191080291227727
%-0.7507129901094797
%[0.00084887 0.01338308 0.02275319]
%[ 8. 10. 12. 14. 16. 20. 24. 32. 40.]
%0.8372(8)L - 0.12(1)\log(L) - 0.75(2)

\section{X. Insensitivity of corner coefficient with $\beta/L$ ratio}
\label{sec:finiteT}

In this section we would like to address a pressing question regarding the extraction of the corner logarithmic contributions at finite temperature.  In principle, the area-law with logarithmic correction scaling form is valid for groundstate wavefunctions of quantum systems.  However, quantum Monte Carlo simulations at 2+1 dimensional critical points typically work at finite temperature, where the inverse temperature $\beta J$ is scaled in proportion to the system size $L$ (as is natural for theories with dynamical critical exponent $z=1$).  So far it is unclear whether the CFT scaling form for $S_2$ continues to hold in this finite-temperature scaling regime.

To check this, and to ensure our results are free from finite-temperature effects, we have performed simulations of the $J$-$Q$ model on the tilted square lattice at the critical point for several different scalings of the inverse temperature $\beta/ L=0.5,1,2$ (in units where $Q=1$).  These results are given in Fig. \ref{fig:cornerbeta}, where we find the extracted corner coefficient for a square subsystem is insensitive to the value of $\beta / L$ that is chosen.  Moreover, we find that the raw $S_2$ data for the square subsystems is essentially converged to the groundstate for $\beta/L=1,2$, which can be seen by the zoom-in in the left panel.  Importantly, even though $\beta/L=0.5$ is clearly not converged to $T=0$, it still results in an extracted corner coefficient that is consistent with the other data sets.  

Out of precaution, we add that one cannot definitively rule out finite-temperature effects based on what we present here, though we believe any possible effects are too small to significantly influence our results.  As a best practice rule of thumb, a similar finite-temperature analysis should be carried out for any related entanglement entropy study.

\end{document}